\documentclass[fleqn,usenatbib]{mnras}

\usepackage{amsmath}
\usepackage{graphicx}

\usepackage[T1]{fontenc}
\usepackage{ae,aecompl}

\usepackage{graphicx}	
\usepackage{amssymb}	
\usepackage{caption}
\usepackage{soul} 

\newcommand{\angstrom}{\mbox{\normalfont\AA}}

\title[Binary stars during reionization]{The effects of binary stars on galaxies and metal-enriched gas during reionization}

\author[C. Doughty et al.]{
Caitlin Doughty,$^{1}$\thanks{E-mail: cdoughty@nmsu.edu}
Kristian Finlator,$^{1,2}$
\\
$^{1}$Department of Astronomy, New Mexico State University, Las Cruces, New Mexico, USA\\
$^{2}$Cosmic Dawn Center (DAWN) at the Niels Bohr Institute, University of Copenhagen and National Space Institute,\\Technical University of Denmark\\
}

\date{Accepted XXX. Received YYY; in original form ZZZ}

\pubyear{2021}

\begin{document}
\label{firstpage}
\pagerange{\pageref{firstpage}--\pageref{lastpage}}
\maketitle

\begin{abstract}
Binary stars are abundant in nearby galaxies, but are typically unaccounted for in simulations of the high redshift Universe. Stellar population synthesis models that include the effects of binary evolution result in greater relative abundances of ionizing photons that could significantly affect the ambient ionizing background during the epoch of hydrogen reionization, additionally leading to differences in galaxy gas content and star formation. We use hydrodynamic cosmological simulations including \textit{in situ} multifrequency radiative transfer to evaluate the effects of a high binary fraction in reionization-era galaxies on traits of the early intergalactic medium and the abundance of H I and He II ionizing photons. We further extend this to analyze the traits of enriched gas. In comparing metrics generated using a fiducial simulation assuming single stars with one incorporating a high binary fraction, we find that binary stars cause H I reionization to complete earlier and at an accelerated pace, while also increasing the abundances of high-ionization metals (C IV and Si IV) in simulated absorption spectra while reducing the abundance of low-ionization states (O I, Si II, and C II). However, through increased photoheating of galactic and circumgalactic gas, they simultaneously reduce the rate of star formation in low-mass galaxies, slowing the ongoing process of enrichment and suppressing their own ionizing background. This potentially contributes to a slower He II reionization process at $z\geq5$, and further indicates that self-regulation of galaxies could be underestimated when neglecting binary stellar evolution.
\end{abstract}



\section{Introduction}
Reionization of hydrogen began with the creation of the first sources of ionizing radiation, but was a protracted process extending over hundreds of millions of years~\citep[see][and references therein for a thorough review]{fan06}. A wide array of strategies have been adopted to study the Universe's evolution in this time. Observations of the red damping wings in $z\sim$ 7 QSOs, Lyman-$\alpha$ Emitter (LAE) detections, and Lyman-$\alpha$ emission from faint Lyman Break Galaxies (LBGs) have provided some constraints on the evolution of the neutral hydrogen fraction during this epoch, placing $x_{HI} \approx 0.5$ at $z\sim8$ and approaching zero by $z\sim6$~\citep{ouchi10, banados18, hoag19}. The appearance of Gunn-Peterson absorption~\citep{gunn65} at $z\sim6$ has also led to the convention that H I reionization concluded at around this time~\citep[see e.g.][]{becker01,becker15,fan06,venemans13}. Measurements of the Thomson optical depth from the cosmic microwave background suggest a reionization redshift of $z_\mathrm{re}= 8.2^{+1.0}_{-1.2}$, assuming it is instantaneous~\citep{planck16-reion-constraints}. However, simulations recreating the observed Lyman-$\beta$ forest opacity at $z=5.5$ support a later conclusion, closer even to $z\sim5.2$~\citep{keating19,kulkarni19}. 

Galaxies have a direct impact on the progression of reionization, but the details of their contributions are unknown. Their intrinsic ionizing emissivity and the escape fraction of ionizing Lyman continuum (LyC) radiation are not directly observable for galaxies at $z>4$ because of the high opacity of the IGM to ionizing radiation. Studies rely on integration of the observed UV luminosity function (UVLF) down to the faint end cutoff in the function, from theoretical work believed to be $M_\mathrm{UV} \sim -13$ ~\citep[e.g.][]{finkelstein15}, and an adopted ionizing photon production efficiency derived from models of stellar populations that make general assumptions about metallicity, binarity, and the star formation history. Faint galaxies are held to represent the largest contribution to the ionizing emissivity, due in significant part to the apparent steepening of the faint-end slope of the UVLF at $z>4$~\citep{bouwens15, finkelstein15}, although the issue is not completely settled~\citep[see e.g.][]{naidu20}. If faint galaxies are indeed the dominant sources of reionization, they may require high average escape fractions of $f_\mathrm{esc} \geq 20$ percent~\citep{finkelstein12}. 

Such high $f_\mathrm{esc}$ values are at odds with the inability of hydrodynamical simulations to recreate values greater than a few per cent~\citep{kimm14,wise14,paardekooper15,ma15,ma16} and additionally the dearth of detections of LyC even for nearby galaxies. While some targets specifically selected for their unusual emission have yielded LyC detections, these galaxies are fairly unique systems and likely not representative of the general population~\citep{steidel01,shapley16,debarros16,bian17,vanzella18,fletcher19,izotov16a,izotov16b,izotov18}. Studies of larger numbers of galaxies without these special emissive characteristics have solidified limits on $f_{esc}$ at $<2$ percent~\citep{siana10,sandberg15,rutkowski17,japelj17,grazian17,hernandez18}, although exceptions have been seen in samples of LBGs~\citep{steidel18}.

The escape fractions generated in simulations are strongly model-dependent~\citep[see e.g.][]{ma15,ma16}, and in particular are affected by assumptions about the stellar populations' metallicity and binarity; EoR galaxies are typically assumed to have relatively low stellar metallicities and consist of single stars. However, observations of nearby dwarf galaxies and star clusters suggest that the fraction of stars in binary systems may be much higher than previously thought~\citep{duchene13a,sana13}. In fact, stars in binary systems are now believed to be quite common, with massive ($>8 M_{\odot}$) O and B types having binary fractions ranging from 60\textendash80+ percent, the fraction decreasing with diminished stellar mass~\citep{duchene13b}.\footnote{Binary fraction here is defined as the fraction of stellar systems that have $\geq 2$ stars.} Binary interactions modify the physical properties of stars and, in doing so, alter the characteristics of the entire population. Changes to their hydrogen envelopes from stripping or mass transfer result in bluer spectra, and differences in mass retention lead to a greater surviving stellar mass throughout the lifetime of a stellar population~\citep{eldridge17}. Overall, several groups have found that bluer spectra result when accounting for binary star system phenomena such as mass transfer and mergers, regardless of the precise details of implementation~\citep{belkus03,zhang15,gotberg19}.

The seeming ubiquity of binary stars in the low-redshift Universe and the UV-brightening effects of interactions on their spectra has lent credence to the possibility of introducing binary stars into cosmological simulations in order to reduce the gap between required and modelled ionizing escape fractions. Works incorporating high binary fractions result in an increase in the ionizing photon budget available to contribute to hydrogen reionization~\citep{ma16, stanway16, rosdahl18}. Because newly forming stars are trapped within high density, ionizing radiation-absorbing molecular clouds, the ionizing output of single stars is stifled when they are at their brightest, at ages $<$ 3 Myr. Longer ionizing lifetimes of binary stars work to counteract this~\citep{ma16}. Further, the discrepancy between single and binary star populations is greater for low metallicities than for solar values, suggesting that it could be even more important to consider binary evolution effects in the reionization era, where typical stars were not yet polluted to solar-level metallicities.

In addition to studies of the escape fractions of galaxies, there are other observables that may be sensitive the presence of high binary fraction stellar populations. Theoretical studies of metal absorption in the circumgalactic and intergalactic media during hydrogen reionization suggest that metal transitions sensitive to the ionizing background may serve as useful probes of the progress and sources of reionization~\citep{oh02,finlator16,garcia17b,doughty18,doughty19,hennawi20}.

In this study, we examine the effects on simulated QSO metal absorption lines of a stellar population synthesis model including binary stars, the Binary Population and Spectral Synthesis model~\citep{eldridge17}, in a cosmological hydrodynamic simulation including \emph{in situ} multi-frequency radiative transfer. While other works have studied this topic~\cite[e.g.][]{ma16,rosdahl18}, we are exploring in more detail the effects of the hardness of the UVB on CGM and IGM heating, on galaxies, and tracers of CGM ionization. In Section~\ref{sec:simulations} we describe the simulations used as well as the stellar population synthesis models, and briefly describe the procedure used to generate mock absorption spectra. Section~\ref{sec:results} describes our results, addressing variations in star formation and galaxy properties, H I and He II reionization, adjustments to typical gas phase and metallicities, and metal absorber properties. We tie the results together and place them within the wider context in Section~\ref{sec:discussion}, and finally present our conclusions in Section~\ref{sec:conclusions}.

\section{Simulations}\label{sec:simulations}
We use an updated version of the {\sc technicolor dawn} (TD) code~\citep{finlator18}, which in turn is built on {\sc gadget}-3 (last described in~\citealt{springel05}). We use a density-independent formulation of smoothed particle hydrodynamics (SPH), enabling the code to model fluid instabilities accurately~\citep{hopkins13}. SPH particles are modeled at $320^3$ resolution in a periodic box of comoving length 12 h$^{-1}$ cMpc (corresponding to a gas particle mass of $M_g = 1.0 \times 10^6 M_{\odot}$), while the UVB is modeled in $32^3$ voxels and 24 independent frequency bins including galaxy and quasar contributions.  The spatially-invariant emissivity of quasars is modeled according to \citet{lusso15} and \citet{manti17}. Gas particles cool radiatively owing mostly to collisional excitation of H, He, and metals; the full tables of processes and rates are taken from~\citet{katz96} and~\citet{sutherland93}. Metal enrichment, including but not limited to carbon, oxygen, silicon and magnesium, is modelled self-consistently in each particle in the simulation, with element quantities generated according to Type Ia ~\citep{heringer17} and Type II ~\citep{nomoto06} supernovae (SNe), and asymptotic giant branch (AGB) yields~\citep{oppenheimer09}.

We model star formation within a multiphase interstellar medium using the prescription of~\citet{springel03}. Gas particles
whose proper hydrogen number densities exceed 0.13 cm$^{-3}$ acquire a subgrid two-phase structure consisting of cold condensed clouds in pressure equilibrium with an ambient hot medium. Star formation proceeds via a Monte Carlo prescription in which the probability that a gas particle spawns a star particle is decided so that the emergent relationship between star formation rate and local gas density matches the observed Kennicutt-Schmidt law~\citep{schmidt59,kennicutt98}. To be consistent with our adopted~\cite{kroupa01} stellar initial mass function, we set the fraction of stellar mass that is returned instantaneously (denoted as $\beta$ in~\citealt{springel03}) equal to 0.187.

There are several changes to the functioning of TD compared to the original description in~\citet{finlator18}, in particular (1) the model for the escape fraction of ionizing photons and (2) the mass-loading factor of galactic winds. As previously, the escape fraction is treated as redshift-dependent and modeled after~\citet{hm12}:
\begin{equation}\label{eq:new_fesc_gal}
f_\mathrm{esc,gal} = 0.166 \left(\frac{1+z}{6}\right)^{2.65}
\end{equation}
A maximum value for the escape fraction, $f_\mathrm{esc,max}=0.31$, is also enforced.\footnote{The factor of 0.166 has decreased from 0.176 and the exponent has settled at 2.65 from values of 1.95, 3.4, and 3.45 in~\citet{finlator18}, where exact values were tuned to ensure that reionization completed by $z=6$ and therefore varied between simulations depending on the resolution of the radiative transfer solver. Previously $f_\mathrm{esc, max}= $ 0.36 and 0.5, again dependent on the RT resolution.} The escape fraction here describes the relationship between the intrinsic ionizing emissivity of the star-forming gas particles and the true total ionizing emissivity in their host RT cell. Effectively, it represents the escape fraction from the unresolved ISM rather than, for example, from a galaxy's virial radius.

The dependence of the mass-loading factor on the galaxy stellar mass has been similarly adjusted. The relation is still based in results from~\citet{muratov15}, and formulated as
\begin{equation}\label{eq:new_mass_loading}
\eta \left(M_*\right) = 2.268 \left( \frac{M_*}{10^{10} M_{\odot}}\right)^{-0.35} 
\end{equation}
but the factor out front has been reduced to 63 per cent of the original value. The justification for these changes is as follows: Although the $2\times256^3$ simulation in~\citet{finlator18} was able to reproduce the UVLF seen in observations~\citep[e.g.][]{bouwens15,finkelstein15,livermore17}, the higher resolution of $2\times512^3$ noticeably underproduced the abundance of UV-bright galaxies. The effect seemed to be caused as gas reservoirs in small systems were severely depleted at early times by efficient star formation and expulsion by winds.  Since $f_\mathrm{esc,gal}$ is one of the few parameters in the simulation that is tuned to reproduce observations, it was adjusted along with the mass-loading~\citep[in accordance with the uncertainties quoted in][]{muratov15} to reduce the discrepancy in the simulated UVLF while still matching the ionizing emissivity at $z=5$ inferred from observations. These alterations were additionally motivated to ensure consistency with measurements of the mean Lyman-$\alpha$ transmission post-reionization~\citep{bosman18}. For other details of the implementation, we refer the reader to~\citet{finlator18} and~\citet{finlator20}.

\subsection{Stellar population models}\label{ssec:stellar_popmods}
Apart from the escape fraction model, the key link between star formation and the metagalactic UVB is the stellar population synthesis model, which determines the ionizing emissivity. As described in~\citet{finlator18}, our fiducial emissivity model is {\sc yggdrasil}~\citep{zackrisson11}, which accounts for low metallicity stellar populations but not the effects of binary stellar evolution. This choice is motivated both by the unique characteristics of metal-poor stellar populations at early times, and in order to facilitate comparison with our previous investigations. We assume a~\citet{kroupa01} IMF, with stellar masses ranging from 0.1 to 100 $M_{\odot}$. Star-forming gas particles are assigned an emissivity based on a 100 Myr constant SFR$=$1 M$_\odot$ yr$^{-1}$ stellar population, re-scaled according to the true SFR of the particle and interpolated to its metallicity. Although {\sc yggdrasil} has the capability of modeling the effects of photoionized gas on the ionizing background, implemented using {\sc cloudy}~\citep{ferland98}, we do not use this feature. In practice, the ions that we consider are only minimally affected by emission from ionized gas because their ionization energies exceed $\sim1$ Ryd, while radiation from photoionized gas (with the exception of He II Lyman-series emission) is generally restricted to redder wavelengths.

As a comparison, we explore the Binary Population And Spectral Synthesis~\citep[{\sc bpass};][]{eldridge17} code, based on Cambridge {\sc Stars}~\citep{eggleton71} which uses a numerical non-Lagrangian mesh to solve for stellar structure while stepping through time from the zero age main sequence to the moment where it ends its evolution. The {\sc bpass} v2.2 data release contains SEDs for both single and binary stars with metal mass fractions spanning $Z=10^{-5}$ to 0.04 ($7\times10^{-4}$ to $2.9 Z_\odot$ assuming $Z_\odot=0.014$) and with various initial mass functions covering subsets of mass from 1$M_{\odot}$ to 300$M_{\odot}$. For the binary SEDs, a flat distribution of initial mass ratios and logged period values is assumed and the distribution of orbital parameters is tuned so that around 80 per cent of the $M > 5 M_{\odot}$ binaries will experience an interaction at some point. In order to have the most compatibility with our previous work using~\citet{kroupa01}, we select an IMF with slope $m=-1.3$ for $0.1 M_{\odot}<M_\mathrm{star}<0.5 M_{\odot}$ and $m=-2.35$ for $0.5 M_{\odot} < M_\mathrm{star} < 100 M_{\odot}$. The instantaneous emissivity is calculated as suggested in the {\sc bpass} manual, according to
\begin{equation}
    F(\lambda) = f_0 \Delta t_0 + \sum_{i=1}^{t_\mathrm{max}} \frac{SFR}{10^6\;M_\odot} f_i (\lambda) \Delta t_i
\end{equation}
\begin{figure}
\centering
\includegraphics[width=0.5\textwidth]{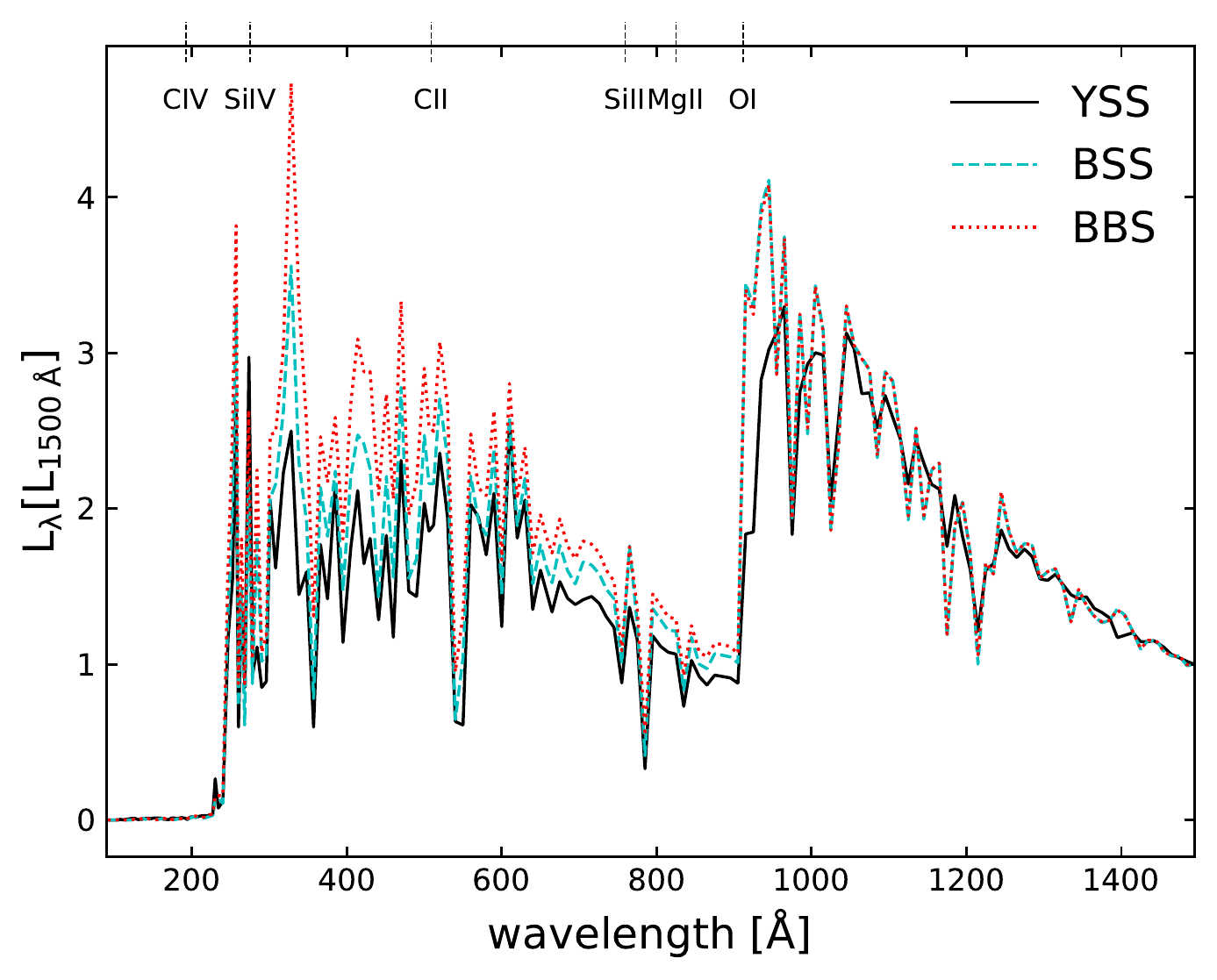}
\vspace{-0.5cm}
\caption{Spectral energy distributions of the {\sc yggdrasil}, {\sc bpass} single, and {\sc bpass} binary stellar population synthesis models assuming a population of $Z=0.004$ forming stars at 1 $M_{\odot}$ yr$^{-1}$ for 100 Myr and normalized to 1.0 at 1500 $\angstrom$. The ionization energies of our considered ions are marked with vertical dashed lines.}\label{fig:SED}
\end{figure}
where $f_i$ is the instantaneous flux at one wavelength when the stellar population is within an age bin $\Delta t_i$.

In Figure~\ref{fig:SED}, we compare the relative luminosities of the three stellar population synthesis configurations used in this work: one with binary stars ({\sc bpass} binary, BBS) and two without ({\sc yggdrasil}, YSS and {\sc bpass} single, BSS). All three are for a metal mass fraction $Z=0.004$ and assume a constant SFR of 1$M_{\odot}$ yr$^{-1}$ for 100 Myr. Broadly, as reported in~\citet{eldridge17}, the BBS model boosts the $\lambda < 912 \angstrom$ luminosity when compared to single stars. 

The fiducial SPS model YSS and the BSS stellar spectra show differences in spectral shape and amplitude. In particular, BSS has a greater amplitude across most of the $\lambda = 90\rightarrow 1500$ range, although the two models produce approximately the same fraction of H I-ionizing photons relative to their total photon production. This is simply because of the differences between the two models in their precise implementations of stellar evolution. The IMFs we adopt are similar, although the high-mass end slope for BSS is steeper than in YSS (-2.35 and -2.3, respectively). While this suggests a slightly higher contribution from massive stars in YSS, it should not be significant.

\begin{figure}
\centering
\includegraphics[width=0.5\textwidth]{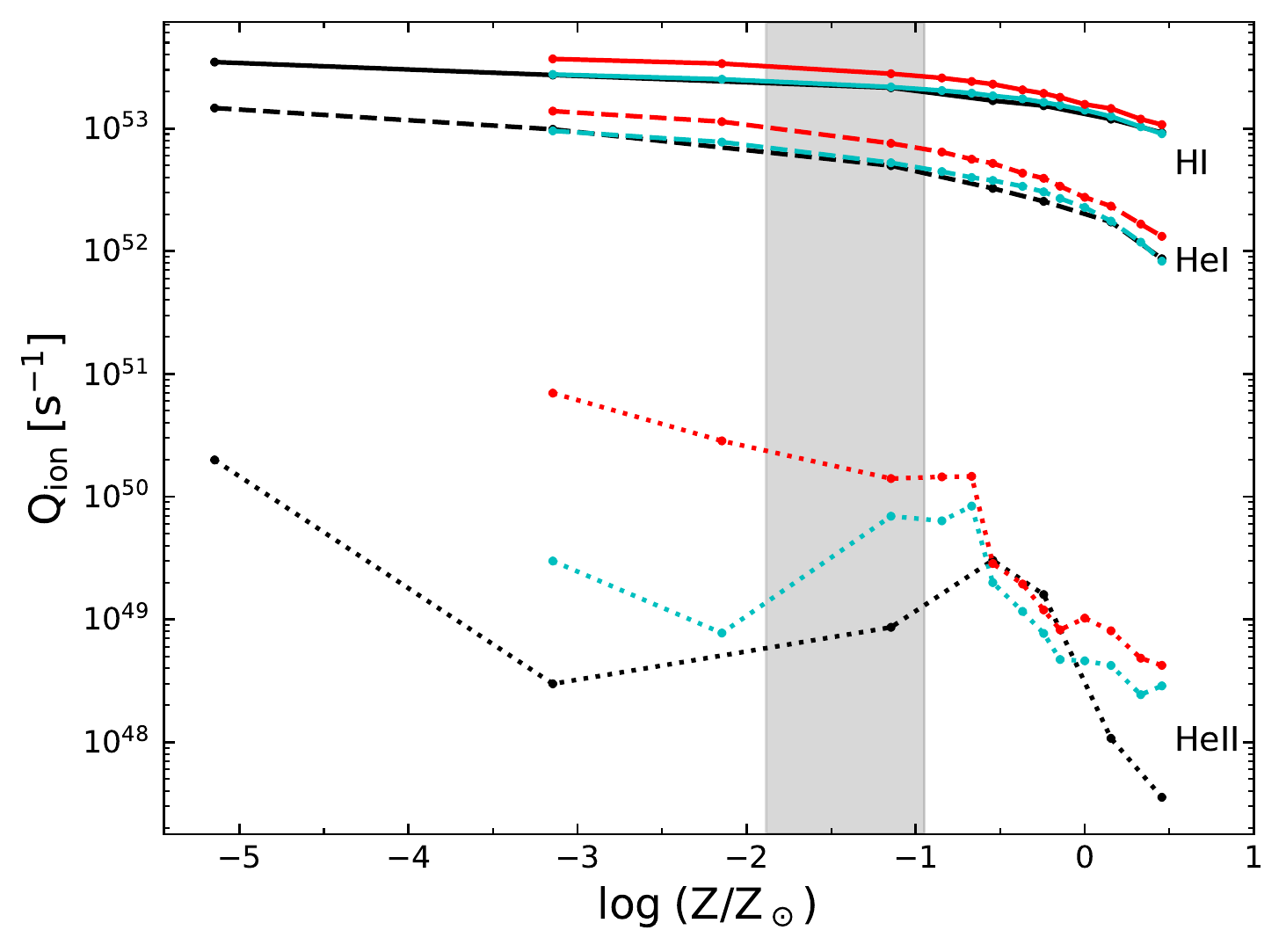}
\vspace{-0.5cm}
\caption{The ionizing photon emissivity for three species by stellar populations that have been forming stars at a rate of 1 $M_{\odot}$ yr$^{-1}$ for 100 Myr as a function of the metal mass fraction. Markers indicate the locations of metallicities used in the SPS models, while the connecting lines show the linearly interpolated values at intermediate Z. The gray shaded region shows the 1 $\sigma$ scatter around the average metallicity at $z=6$.}\label{fig:ionizing_photon_emissivity}
\end{figure}

In Figure~\ref{fig:ionizing_photon_emissivity}, we show the ionizing photon emissivity for the H I, He I, and He II transitions generated by each model SED as a function of metallicity. This is given by
\begin{equation}
    Q_\mathrm{ion} = \int^{\infty}_{\nu_\mathrm{ion}} \frac{L_{\nu}}{h \nu} d \nu
\end{equation}
where $L_\mathrm{\nu}$ is the luminosity at some frequency $\nu$ (from the same type of stellar population as in Figure~\ref{fig:SED}), and $\nu_\mathrm{ion}$ is the transition frequency for a given species. As in Figure~\ref{fig:SED}, binary stars show greater relative emissivity at shorter wavelengths, and the effect is larger for lower metal mass fractions. For example, for metal mass fractions of $\log (Z/Z_\odot)=-0.5$, the binaries outproduce H I-ionizing photons compared to BSS by 25 per cent and YSS singles by 36 percent. At 1 dex higher in metallicity, $(Z/Z_\odot)= 0.5$, 
the BBS SED emits 19 per cent more than the BSS model and 16 per cent more than YSS.

For the two {\sc bpass} models, the predicted rate of ionizing photon emissivity is consistently higher when including binary stars, regardless of metallicity or the particular transition. The harder nature of binary spectra is also visible, with the emissivity for the He I transition at $\lambda = 504.6 \angstrom$ displaying a larger inter-model difference than for H I. Comparing the BSS and YSS models, the H I-ionizing emissivities are nearly indistinguishable, as are the He I-ionizing emissivities. However, for He II especially at lower $Z$, they are consistently larger for the BSS model. In summary then, we expect binary evolution to boost the stellar ionizing emissivity in the simulation at energies above the Lyman limit.

There is a critical point to emphasize before proceeding, which is that the SPS models are the \emph{only} varied parameter between the three simulations used in this work. The escape fraction model, while physically motivated, has been tuned through previous experimentation such that the fiducial YSS simulation completes reionization at $z\sim6$, in accordance with observations. Since the escape fraction model is held constant, variation in the timing of reionization that results from the addition of binary stars is expected due to their increased production of ionizing photons. However, because of the specific tuning of the escape fraction model, the comparison between observations and simulations conducted in Sections~\ref{sec:results} can't be used to rule out the presence of binary stars during reionization; we can only draw conclusions about their effects relative to those of single stars and rule in/out binary stars in combination with our fiducial escape fraction model.

\subsection{Simulated metal line absorption systems}\label{ssec:sim_metal_lines}
To generate absorption systems, a sightline is cast through the simulation volume, wrapping periodically through the volume oblique to the boundaries to ensure that the entire 
\begin{figure}
\includegraphics[width=0.5\textwidth]{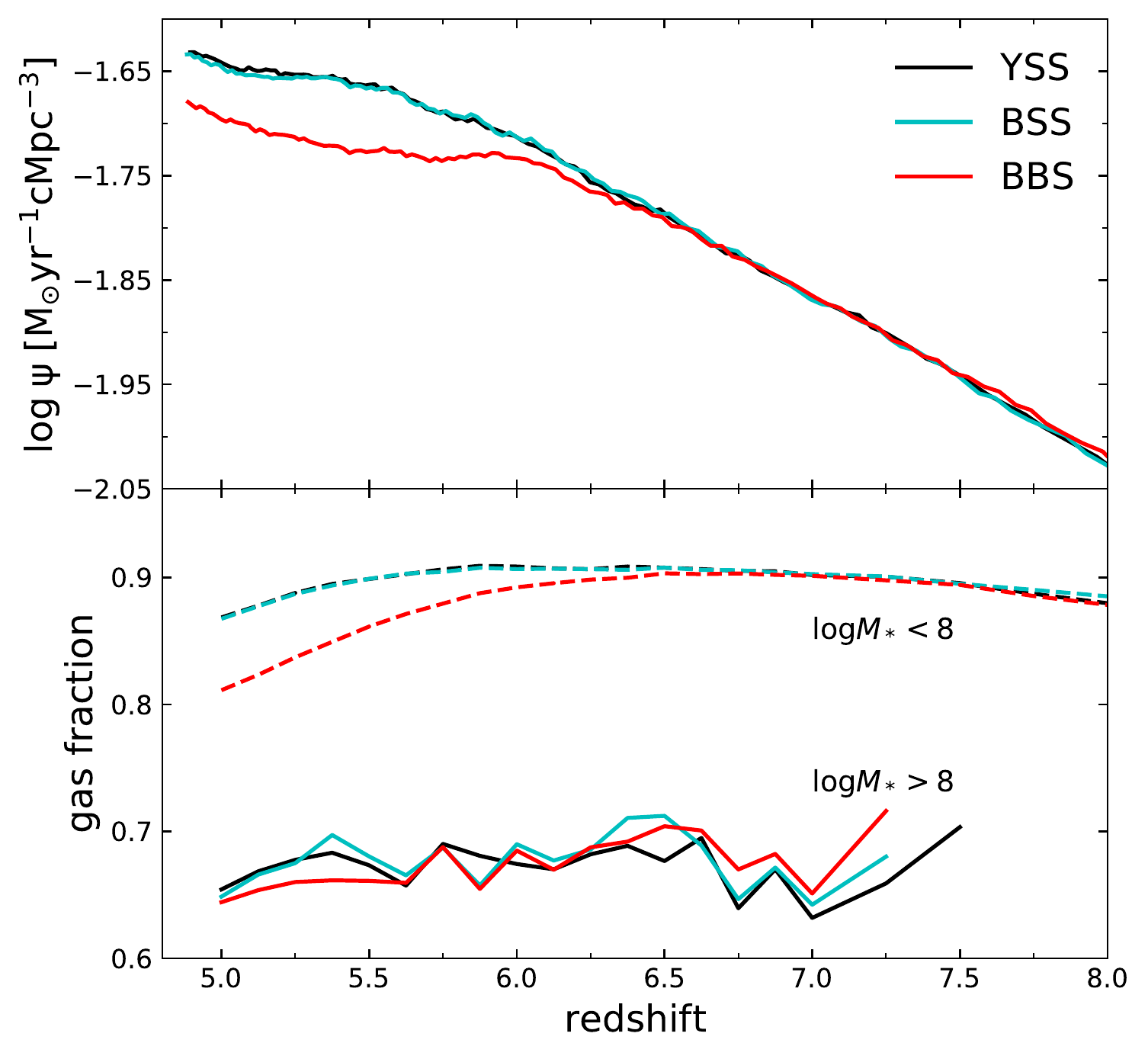}
\vspace{-0.5cm}
\caption{\emph{Upper panel:} The total star formation rate output from the simulation. BBS shows decreased star formation by $z=6$ and has fallen to $\sim 90$ per cent of the single star models by $z=5$. \emph{Lower panel:} The gas content of halos as a fraction of their stellar mass, divided into halos with $M_* > 10^{8} M_\odot$ (solid) and $M_* < 10^8 M_\odot$ (dashed). The gas fraction in the lower mass bin is consistently higher, and the fraction for the BBS model begins to deviate from the single star models as early as $z=6.5$. } \label{fig:sfrrate}
\end{figure}
volume is well-sampled. The sightline spans $4\times10^6$ km/s in Hubble velocity, and physical properties such as density, metallicity, and temperature of gas particles near the sightline are mapped onto it. When calculating the equilibrium ionization state of each element in post-processing, the local ionizing background is taken into account to apply photo- and collisional ionization and direct, dielectric, and charge transfer recombination. Voigt profiles are used to model the shape of the absorption lines~\citep{humlicek79}, noise appropriate for S/N=50 is added, and the equivalent width is measured directly from the resultant spectra. Absorption systems are identified where the mean transmitted flux drops at least 5$\sigma$ below the continuum for three contiguous pixels. The column densities used in the generated absorber catalog are then calculated using the apparent optical depth method~\citep{savage91}. We consider six particular transitions: O I $\mathrm{\lambda} 1302 \angstrom$, Mg II $\mathrm{\lambda \lambda} 2796 \angstrom$, Si II $\mathrm{\lambda} 1260 \angstrom$, C II $\mathrm{\lambda} 1335 \angstrom$, Si IV $\mathrm{\lambda \lambda} 1394 \angstrom$, and C IV $\mathrm{\lambda \lambda} 1548 \angstrom$, only considering the blueward transition in the case of doublets. In order to be concise, we will exclude the associated wavelengths when describing these transitions, e.g. we will refer to O I $\mathrm{\lambda} 1302 \angstrom$ as O I.

\section{Results}\label{sec:results}
We aim to consider three questions given the scenario of high binary fractions during the epoch of reionization (EoR);
\begin{figure}
\centering
\includegraphics[width=0.5\textwidth]{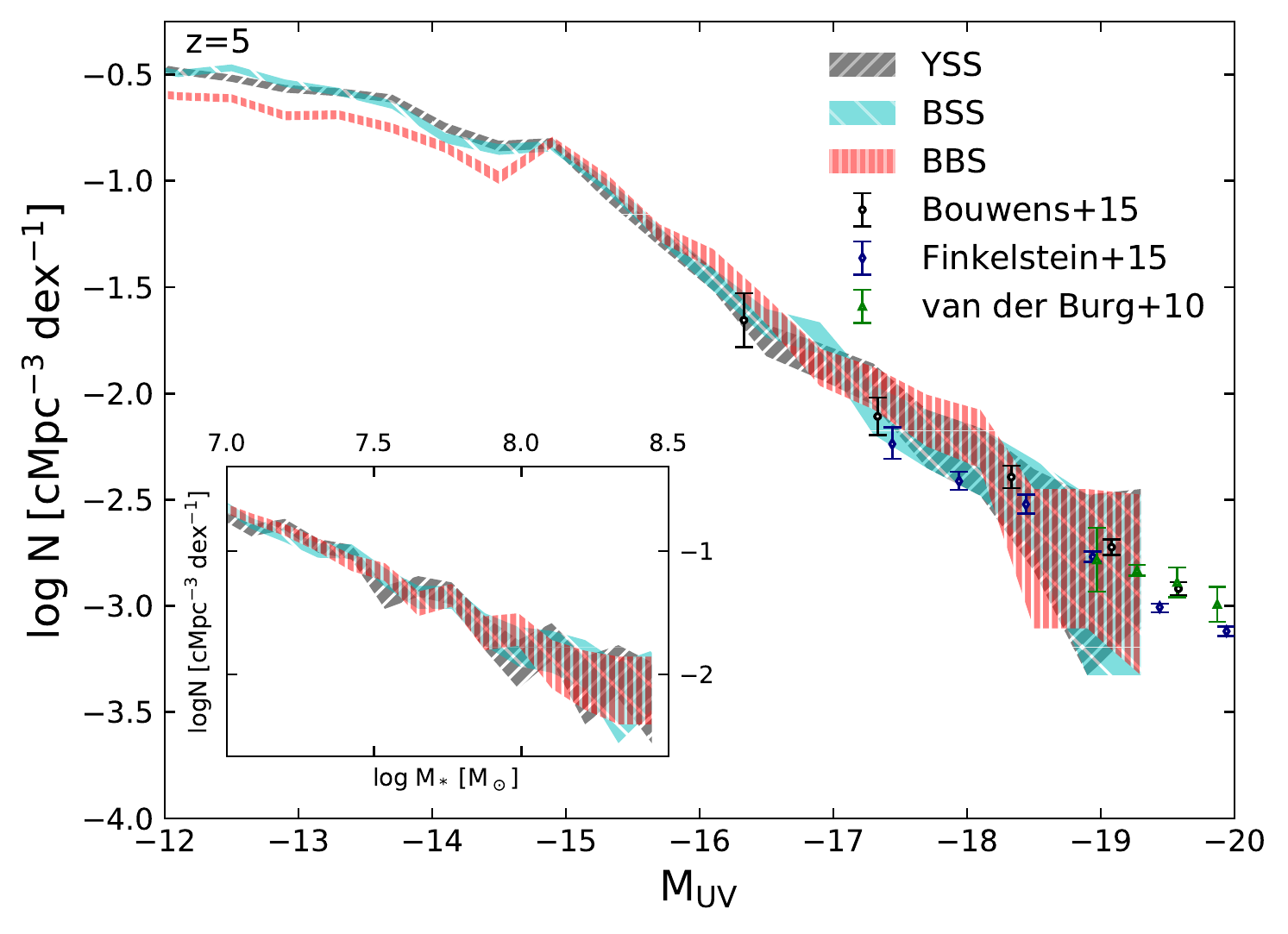}\caption{The UVLF taken at 1500 $\angstrom$ at $z=5$, with overplotted cosmology-corrected observations from~\citet{bouwens15},~\citet{finkelstein15}, and~\citet{vhe10}. The inset plot shows the stellar mass function also at $z=5$, which shows no difference between the models for $\log M_* > 7$.}\label{fig:uvlf}
\end{figure}
\begin{enumerate}
\item How are star formation and galaxy properties altered by the presence of binary stars;
\item What general trends can we expect in the H I and He II reionization histories;
\item How do binaries affect the temperature and metal content/ionization state of the gas around galaxies?
\end{enumerate}

\subsection{Star formation and galaxy stellar properties}
To address our first question pertaining to star formation and galaxy properties, we evaluate the star formation rate density (SFRD). The SFRD (top panel of Figure~\ref{fig:sfrrate}) is essentially the same in all three models for $z>6.5$, barring minor elevation for BBS at $z>7.5$. For lower redshifts though, BBS begins to create fewer stars especially below $z=6$, and by $z=5$ is forming 90 per cent as many stars as the single star models. The changes to the star formation rate do not translate to any obvious differences in the stellar mass function for stellar masses above our resolution limit (see inset of Figure~\ref{fig:uvlf}); however, the calculated $z=5$ UVLF shows a corresponding decrease in the number of galaxies with UV magnitudes $M_\mathrm{UV}>-15$ when binary stars are included, amounting to 20 per cent fewer galaxies between $-15 < M_\mathrm{UV} < -12$. There are no differences at higher redshifts.

We find a potential cause of the drop in SFRD through analysis of the galaxy gas fractions. While for halos with $\log M_* > 8$ the mean fractions in all models are similar for $z<7$, for lower stellar masses there is a downturn at $z\sim6.5$, leading to a relative deficit of $\sim$ 7 per cent by $z=5$ (bottom panel of Figure~\ref{fig:sfrrate}). The suppressed gas fractions in low-mass galaxies almost certainly drive the decreased SFRD. A caveat on this result is that galaxies require at least 64 star particles to be reasonably converged~\citep{dave06}. For our SPH mass resolution, this places the lower stellar mass limit at $\log M_* \approx 7.5$, meaning there may be convergence issues for the lower mass galaxies in the $\log M_* < 8$ bin in Figure 3. However, we expect this result to be qualitatively robust.
\begin{figure}
\includegraphics[width=0.5\textwidth]{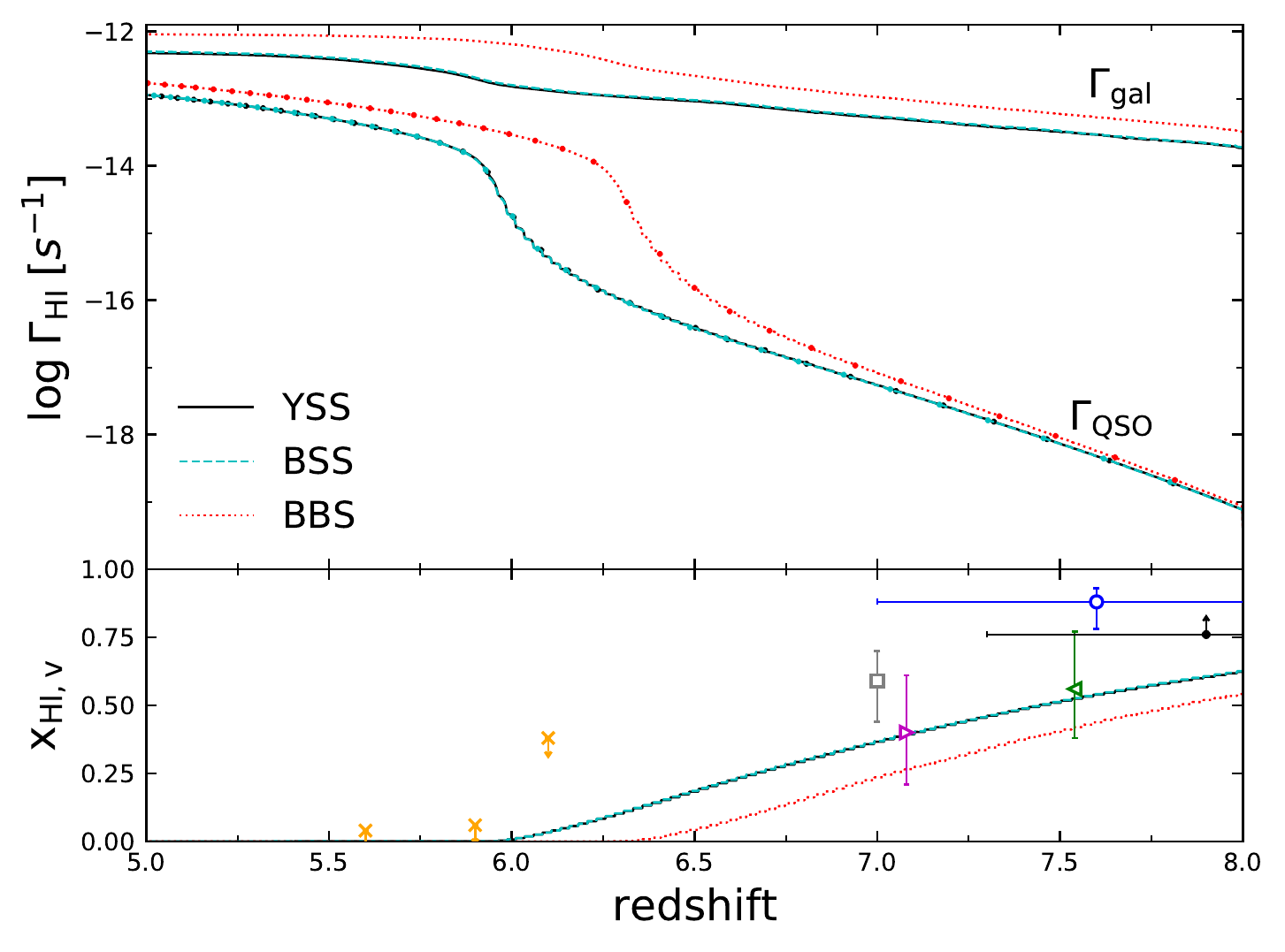}
\vspace{-0.5cm}
\caption{\textit{Upper panel:} The volume-weighted photoionization rate of neutral hydrogen separated into galaxy (no markers) and quasar (with dots) contributions. \textit{Lower panel:} The volume-weighted neutral hydrogen fraction. Observations are shown from~\citet{hoag19} (blue circle),~\citet{banados18} (green leftward triangle),~\citet{greig17} (magenta rightward triangle),~\citet{mcgreer15} (orange exes), and~\citet{mason18,mason19} (gray square and black point, respectively).} \label{fig:xHIv_gammas}
\end{figure}

\subsection{Reionization}
Here we examine how the particular SPS implementation affects the abundance of ionizing photons and the consequences for hydrogen and He II reionization. We begin with the volume-averaged H I photoionization rate, $\Gamma_\mathrm{HI}$, which is the rate of photoionizations occurring per second due to photons generated by stars and quasars:
\begin{equation}\label{eq:avg_photoionization}
    \langle \Gamma_\mathrm{HI} \rangle = \int \frac{4 \pi \langle J_\nu\rangle}{h \nu} \sigma_\mathrm{HI}\left(\nu\right) d\nu 
\end{equation}
where $\sigma_\mathrm{HI}$ is the photoionization cross-section of hydrogen and $\langle J_\nu \rangle$ is the volume-averaged radiation intensity, the latter dependent on both the sources of radiation and the opacity of the IGM. The upper panel of Figure~\ref{fig:xHIv_gammas} shows the redshift evolution of $\Gamma_\mathrm{HI}$, with contributions from galaxies and quasars indicated separately; the lower panel shows the redshift evolution of the volume-weighted neutral hydrogen fraction. 

Regardless of the specific stellar evolution implementations in YSS and BSS, there is no difference in any detail of the H I photoionization rate or reionization history, and likewise no changes in the expectation of contributions from galaxies and quasars, indicating that there are no effective differences in either the galaxy emissivities or IGM opacities at the energies relevant for H I. BBS, on the other hand, creates a boost in $\Gamma_\mathrm{HI, gal}$ compared to single stars. This enhancement is established by $z=8$ and stays relatively constant until $z<6.25$ where we see a jump in photoionization and, by $z=5$, BBS is generating $\log \Gamma_\mathrm{HI,gal} = -12.0$ versus $-12.3$ for BSS. These differences drive an accelerated pace in the reionization history of hydrogen: At $z=8$, the volume-weighted neutral hydrogen fractions are about 10 per cent lower for binary stars and the separation increases until the EoR is complete. BSS reaches $x_\mathrm{HI,v}=0.5$ at $z=7.4$ while BBS achieves the same fraction about 50 million years 
\begin{figure}
\includegraphics[width=0.5\textwidth]{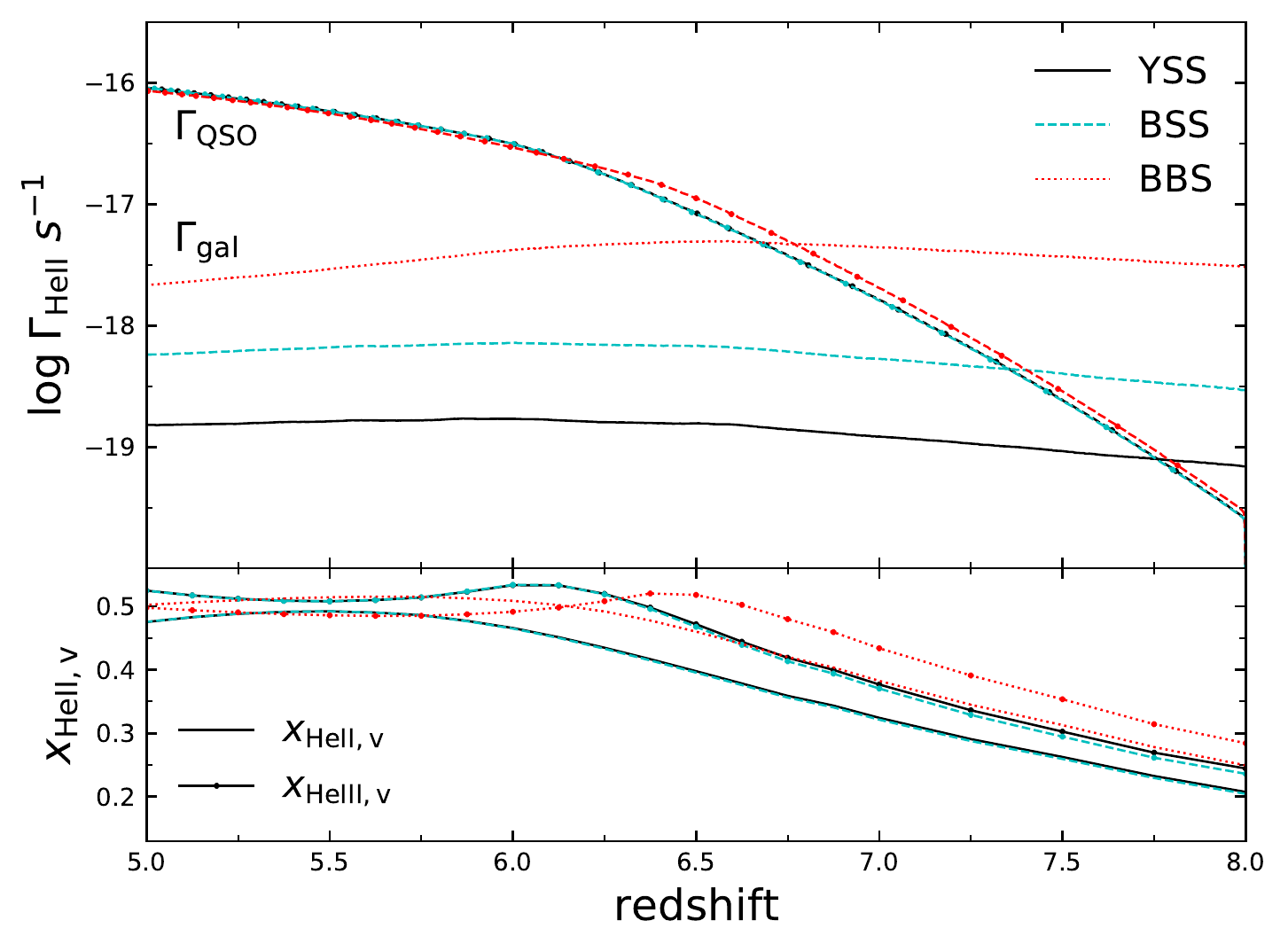}
\vspace{-0.5cm}
\caption{\textit{Upper panel:} The volume-weighted photoionization rate of He II, separated into galaxy (no markers) and quasar (with dots) contributions. \textit{Lower panel:} The volume-weighted He II and He III fractions.}\label{fig:xHeIIv_gammas}
\end{figure}
earlier at $z=7.8$. Decreasing in redshift, the delay between equivalent stages of reionization is enhanced to $\sim$75 million years as BBS achieves the epoch of overlap at $z=6.4$ and BSS by $z=6.0$.

Although galaxies are believed to dominate the reionization of hydrogen, quasars also have some contribution. At $z=8.0$, $\Gamma_\mathrm{HI, QSO}$ is effectively identical for all models, but the BBS value drifts slowly higher with decreasing redshift and undergoes a strong increase below $z=6.5$ compared to BSS and YSS.\footnote{Since only the SPS model inputs differ between the three simulations, one may ask why there is any change in the contribution of quasars to H I photoionization. This is because while the emissivity of quasars is the same, the mean free path is increased due to decreasing IGM opacity in response to the higher ionizing photon budget in the binary star simulation. In radiative equilibrium, this leads to an increase in $\Gamma_\mathrm{HI}$.} A similar, though more gradual, increase occurs for BSS and YSS below $z=6$. The rates begin to converge by $z=5$ as the IGM opacity converges. Ultimately, however, quasars still make up only a minuscule amount of the ionizing photon budget for H I in the simulation due to their reduced number density at EoR redshifts.

Measurements and inferences of the H I neutral fraction have been determined using a wide variety of techniques~\citep{mcgreer15,greig17,banados18,mason18,hoag19,mason19}. The majority of the observations are in agreement with the predicted reionization histories from all three simulations, and those that are not in agreement similarly disagree with all three SPS model results. This comparison is included mostly for illustrative purposes as the early end to reionization in BBS is in part due to the escape fraction model having been previously tuned to accommodate the use of {\sc yggdrasil}.

\begin{figure*}
\includegraphics[width=1.0\textwidth]{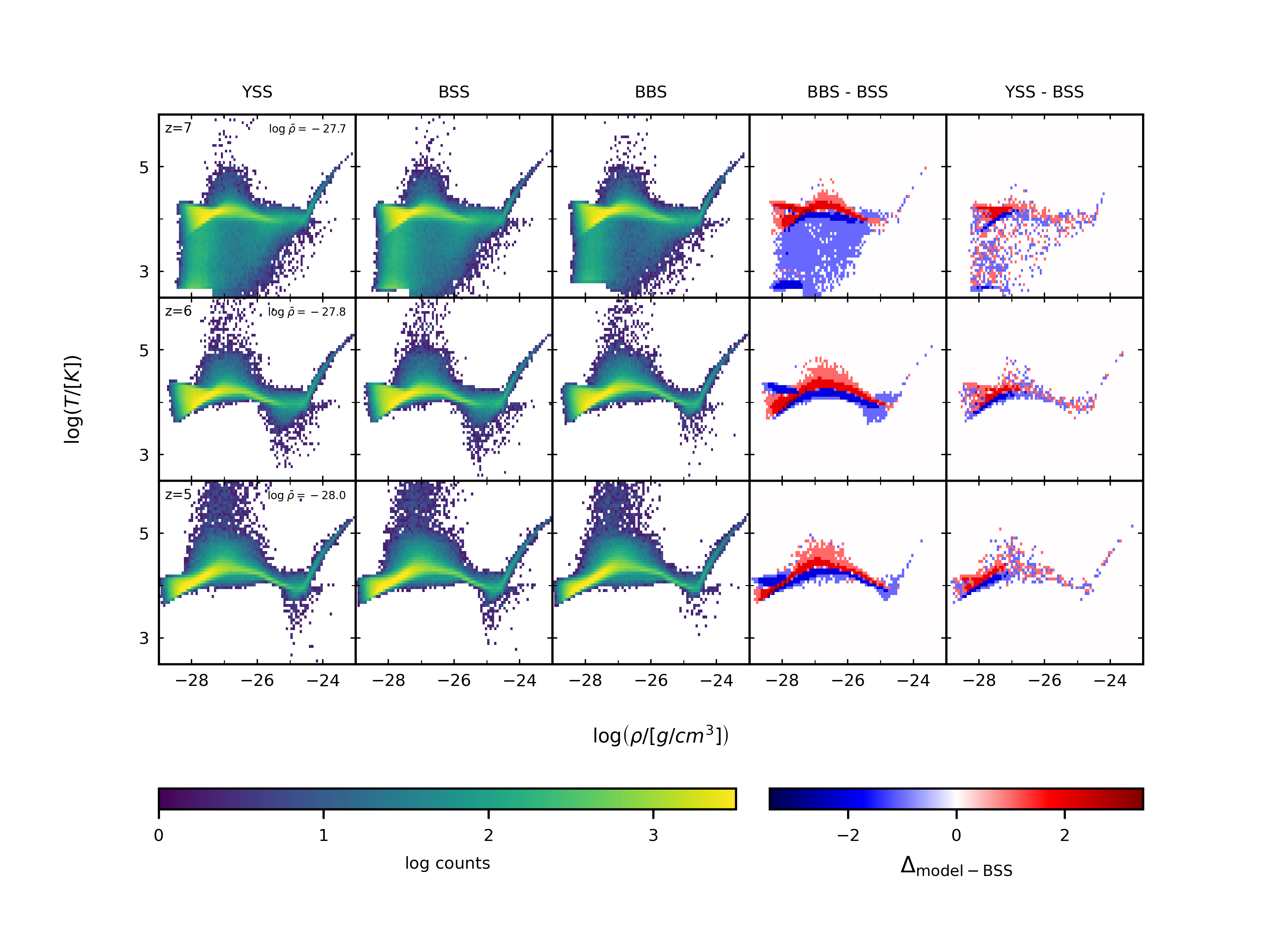}
\vspace{-1.5cm}
\caption{$\rho$-T diagrams for all gas for the three SPS models at three redshifts. The right two columns indicate the residuals between YSS or BBS and the single stars {\sc bpass} model. Positive increases in gas abundance are indicated by red while losses are in blue, with a more saturated color indicating a larger difference. Binary stars result in hotter gas, but there are also some differences between YSS and BSS.} \label{fig:rhot_all}
\end{figure*}

Turning to He II reionization, the photoionization rates for He II from quasars are very similar for both single and binary star models, being slightly elevated for BBS at $z>6$ (Figure~\ref{fig:xHeIIv_gammas}). The most significant difference is seen in galaxy contributions, with differences of approximately 0.5 dex or higher between models, with YSS being the lowest and BBS the highest. Although these changes do not necessarily result in large variations in the He II reionization history, they do affect the redshift at which quasars begin to dominate the photon budget, occurring up to 80 million years earlier for BSS versus BBS ($z=7.3$ as opposed to $z=6.7$).

As with H I reionization, we find no differences in the evolution of the He II and He III fractions, $x_\mathrm{HeII,v}$ and $x_\mathrm{HeIII,v}$, respectively, between YSS and BSS; at $z=8$, single stars create ionization fractions of 20 and 24 per cent for YSS and BSS, respectively, and reach 47 and 52 per cent by $z=5$. The binary stars produce higher fractions of He II and He III at earlier times, with 25 and 28 per cent at $z=8$ and $\sim$50 per cent in both by $z=5$. These differences are not extreme: both single and binary stars seem to result in He II reionization being roughly halfway complete by $z=5$. The slight decrease in the fraction of He III seen for binary stars at later times may be a result of the decrease in star formation also induced by binary stars as seen in the SFRD and the dip in the UVLF for $M_\mathrm{UV}>-15$, as the reduced star formation results in lower overall stellar masses and reduced He II-ionizing photon production.

\subsection{Gas phase}\label{ssec:gas_phase}
Given that binary stars result in a diminished gas fraction in low mass galaxies and consequently a suppression of star formation, it is of interest to explore the impact on the thermal evolution of the IGM. To examine this, we plot $\log \; T$  versus $\log \; \rho$ for a randomly selected subset of all gas in Figure~\ref{fig:rhot_all}, including gas at all overdensities, temperatures, and levels of metal enrichment. By sheer numbers this is dominated by gas in the IGM phase. To highlight the changes between models, we plot the differences in point density in the last two columns in the figure. The IGM, CGM, and ISM are relatively easy to distinguish here. The IGM is at the far left and shows the expected $\rho-T$ power-law relationship starting to form by $z=7$ and becoming more tight by $z=5$. The CGM and ISM show an inversion of this relationship, where increasing density is anti-correlated with temperature. The narrow spike at the right of each panel is created by star-forming gas, where the proper hydrogen number density exceeds $n_H =$ 0.13 cm$^{-3}$.

At $z=7$, all gas is fairly similarly distributed regardless of SPS model, at least by eye. The residuals between the single and binary models, however, show an increase in temperature along the main branch of the gas distribution near $\log \; T \sim 4.0$ and spanning $-28.5 < \log \rho < -26$. This 
\begin{figure}
\includegraphics[width=0.5\textwidth]{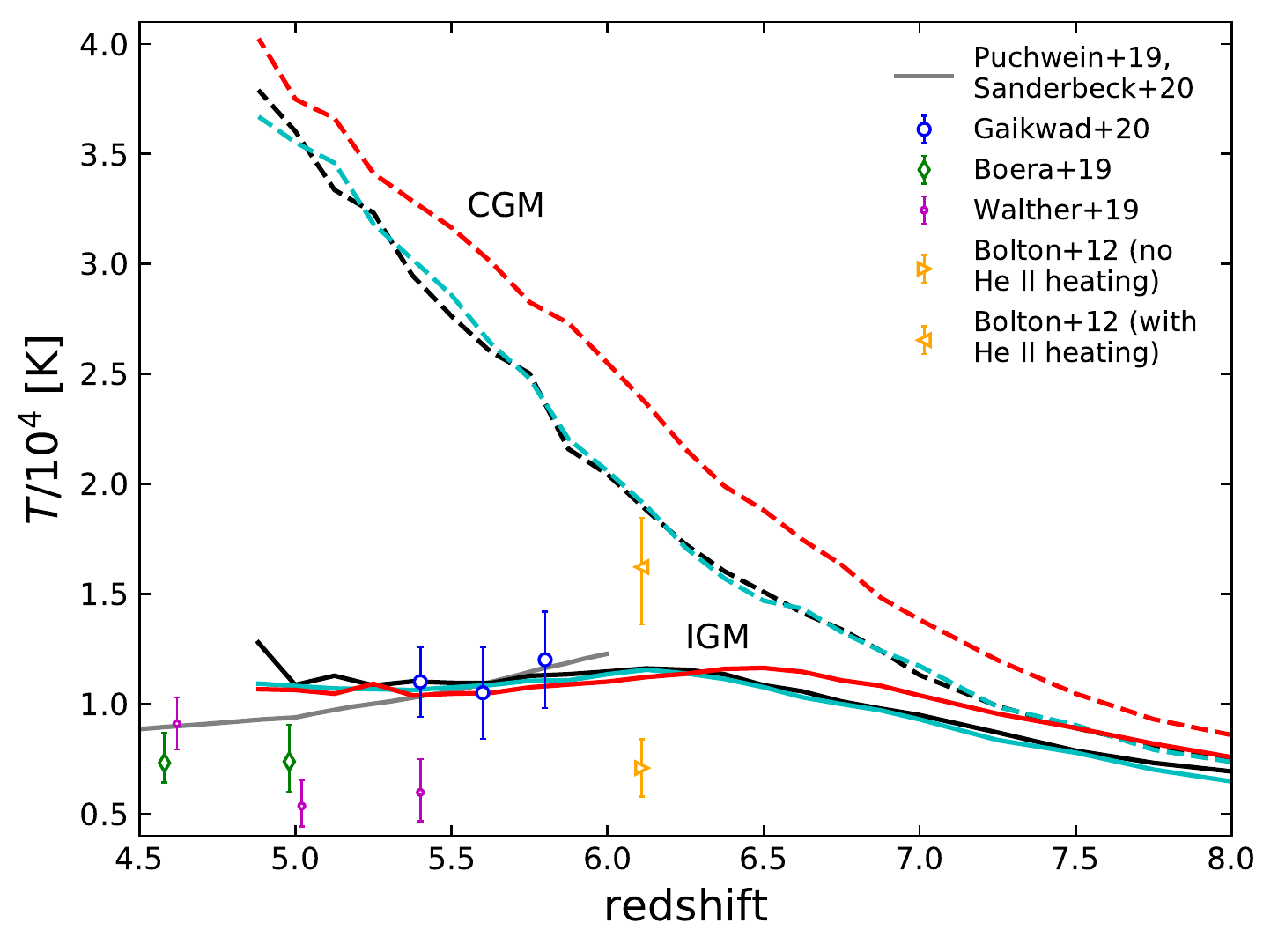}
\vspace{-0.5cm}
\caption{The temperature at mean density (solid) in the simulation as a function of redshift for the three SPS models, calculated as the mean temperature in a bin spanning $\bar{\rho} \pm 0.01 \bar{\rho}$. The mean temperature of the CGM, defined as gas with $1.0 \leq \log \Delta < 2.5$~\citep{pallottini14}, is overplotted. }\label{fig:temp_mean_density}
\end{figure}
adjustment causes the branch to shift upward in $\log \; T$, increasing by over 1000 K. There is also a loss of $\log \; T = 2.8$ and $\log \rho \sim -28.5$ gas in BBS compared to the single star model, visible in the lower left of the $z=7$ residual plots. The YSS model deviates a bit more from BSS here than some previous metrics would indicate, and raises the average temperature of the gas by a few hundred K. Descending to lower redshifts the same general trend is retained, and the low temperature and density variation disappears due to a removal of gas from this phase in all models at later times.

The changes can in part be explained by the stage of H I reionization achieved at each redshift in the three models, as this is strongly associated with the IGM gas temperature. Since by $z=7$ BBS has progressed further in H I reionization than the other two models, we would therefore expect the gas to have reached a greater temperature by a given redshift, even if other details of reionization weren't changed, such as the speed or topology of the process. Also BBS, having completed reionization sooner than the other models, collapses earlier to the expected power-law relation between density and temperature in the post-reionization IGM~\citep{hui97}.

The variation of the gas temperature in response to the changes in stellar populations suggests potential change in the temperature at mean density of the IGM, $T_0$. We calculate this as the mean temperature within a bin in proper density centered on the mean proper density $\bar{\rho}$, where $\bar{\rho} = \Omega_b \rho_\mathrm{crit,0} \left(1+z\right)^3$ (Figure~\ref{fig:temp_mean_density}). Reflecting the changes in the $\rho-T$ distributions, $T_0$ in BBS is elevated by a $\sim$1000 K at early times over the single star models. The characteristic peak in $T_0$ occurring near the completion of H I reionization is earlier for BBS, at $z\sim6.5$ as opposed to $z\sim6$ for both of the single star models, but the value of the peak temperature is roughly the same. BSS tends to have the lowest $T_0$ across all redshifts of the three models, but only by a few hundred K.

Overplotting observationally-derived inferences of $T_0$ near the end of reionization, it is apparent that there are significant differences between several of them and the simulated values. However, there are also conflicts between the observations themselves. $T_0$ as inferred from Ly$\alpha$ transmission spikes~\citep{gaikwad20} spans temperatures of $\sim 1.05 \times 10^4$ to $1.2 \times 10^4$ K at redshifts immediately following the close of reionization, and is in agreement with the temperatures predicted by all three models, although they are lower than the maximum likelihood value. Also in agreement with~\citet{gaikwad20}, a recent analytical metagalactic UVB model from~\citet{puchwein19}~\citep[also used in][]{upton_sanderbeck20}, predicts a much more steeply increasing temperature, and generates higher temperatures than TD for $z>5.4$. However, additional observations using the Ly-$\alpha$ flux power spectrum consistently show lower $T_0$ than other works~\citep{boera19,walther19,bolton12}. Compared to the observations, it seems that the differences in reionization's timing between the single and binary star models do not result in either increased or decreased agreement with the observations.

Turning to higher density gas, we calculate the mean temperature of the CGM. The CGM, here selected to range from 10 to $\sim$316 times the mean density at a given redshift, has an average temperature lower than $T_0$ at very high redshifts ($>7.5$), but increases monotonically with decreasing redshift. This presents a noticeably different trend from that of $T_0$, which begins to decrease once H I reionization is complete. The discrepancy in CGM temperatures is also greater between models than for the IGM, in BBS being about 1200 K greater at $z=8$ and reaching about 3500 K greater by $z\sim5$. Differentiating between the dense and diffuse CGM, the mass-weighted CGM temperature is greater than when volume-weighted, on average by $\sim$1150 K in $6 < z < 8$ for all models. However, the difference between them dissipates by $z=5$. It seems then that the additional heating in BBS has a greater effect on the denser CGM at early times. Overall, the harder background of the binary stars appears to be redistributing all gas to higher temperatures. Although it affects all gas, the largest change in typical temperature occurs for the CGM rather than the IGM.

\subsection{Metal-enriched gas}
Based on Figure~\ref{fig:temp_mean_density}, it is apparent that the CGM is more sensitive to the UVB spectral slope than the IGM, thus it is relevant to quantify the effects of a binary star background on CGM observables. At very high redshifts, while it is not straightforward to derive estimates of the CGM temperature, it is quite possible to observe metal absorption signatures in quasar spectra. Here we examine the changes in overall enrichment, the phase of enriched gas, and statistics of metal absorbers and their variance with SPS model, with the intention of exploring how the bluer spectra of binary stars might be expected to affect the ionization state of metals in the CGM at $z>5$.

\subsubsection{Gas phase and enrichment}
\begin{figure*}
\includegraphics[width=1.0\textwidth]{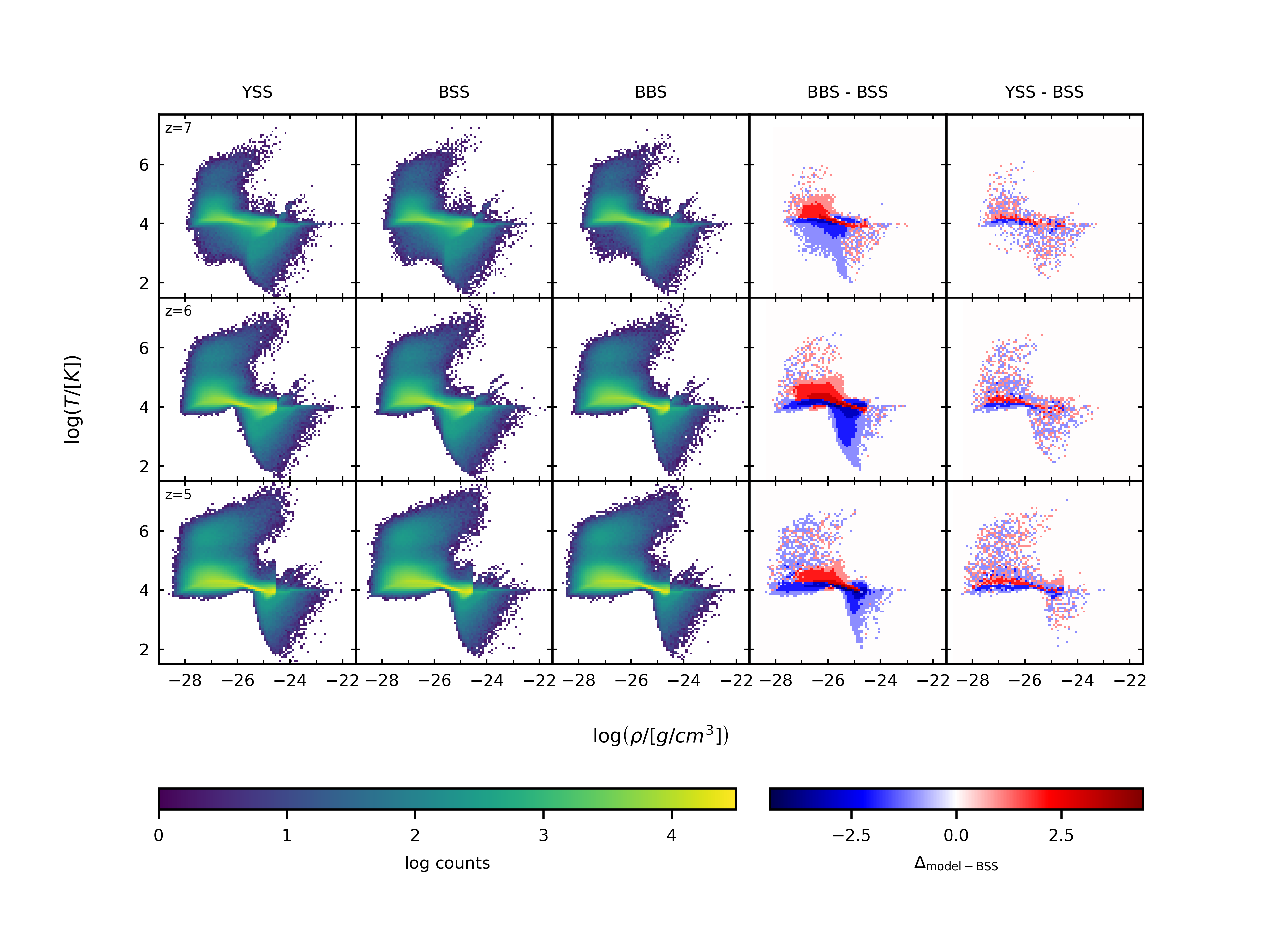}
\vspace{-1.5cm}
\caption{The same as Figure~\ref{fig:rhot_all}, but for gas that is enriched with metals. Whereas for all gas the difference was primarily in temperature, here binary stars remove gas from higher density regions as well.} \label{fig:rhot_enriched}
\end{figure*}

Compared to the distribution of all gas, a larger fraction of the enriched gas is located in two regions of  the $\rho-T$ diagram (Figure~\ref{fig:rhot_enriched}), specifically in more diffuse, hotter regions and denser, cooler regions. 
Between models, the distributions look similar at $z=7$, although by $z=6$ the high-$\rho$ and low-$T$ cloud is wider in the single star models than  in the binary star model. Those visual differences become less apparent by $z=5$. In the residuals, at $z=7$ the main deviation appears to once again be the temperature of the main branch of gas in phase space although the distribution is much less linear than it was for all gas. By including binary stars, there are induced losses in cooler, denser gas and from the main branch, and gains at higher temperatures and lower densities. This trend persists to $z=5$, although the magnitude of the difference does not appear to change much. Some of the difference between models at $z=5$ is due to variation in the amounts and distribution of metals, as BBS experiences a reduction in the total amount of enriched SPH particles; however, this effect is certainly sub-dominant in comparison to the changes in the physical conditions of the gas.

\begin{figure*}
\includegraphics[width=\textwidth]{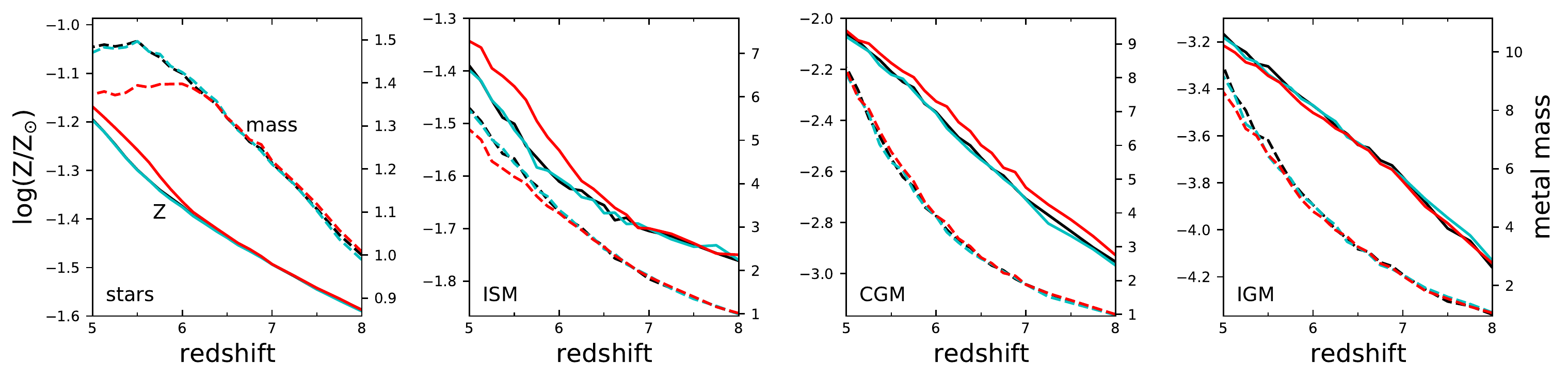}
\vspace{-0.5cm}
\caption{Mass-weighted metallicities (solid, left y-axis values) and total metal masses (dashed, right y-axis values) in several phases of gas, and stars. The metal masses are measured with respect to the mass in YSS at $z=8$ for each phase.}\label{fig:gas_star_Z}
\end{figure*}

In Figure~$\ref{fig:gas_star_Z}$ we show total metal masses and mass-weighted metallicities of the ISM, CGM, and IGM as well as stars as a function of redshift. In the binary star model, the total mass of metals in the stars and the ISM undergoes a decrease relative to single star models at $z\sim6$. The IGM metal mass is slightly reduced in the binary star simulation, being 6 per cent lower than in BSS by $z=5$. Because of the timing, the changes in metal mass of these phases are probably directly associated with the simultaneous downturn in the SFRD in the binary star model. For the BBS CGM, there are slight relative elevations in metal mass for several periods in $z=8\rightarrow$5, but they are not sustained. In both stars and the ISM, the mass-weighted metallicity becomes higher for BBS, below $z=6$ for the stars, but as early as $z=6.7$ for the ISM. The metallicity of the BBS CGM is already elevated by $z=8$, and we find this trend is slightly more prominent in the mass-weighted metallicity than when volume-weighted. The IGM metallicity is slightly reduced at later times by the binary stars. 

Why are the ISM and CGM metallicities noticeably increased when the mass of metals in these phases is not? Investigating this further, we find that the total mass of the BBS ISM is 5 per cent lower than in BSS by $z=6.5$, which increases to a 15 per cent deficit by $z=6$ and 23 per cent by $z=5$. Similarly, the mass of the CGM is lower by $\sim$ 7 per cent in the binary star simulation at $z=8$, which increases to 12 per cent by $z=7$; in complement, the total IGM mass is increased. This is likely to be associated with the temperature increase of the CGM caused by binary stars, which is established well before  $z=8$, as the higher temperatures will cause the density of the CGM gas to decrease and fall within our defined IGM overdensity range of $\log \Delta < 1.0$. A similar effect is possibly happening with the ISM, although it takes longer to become established. 
Generally then, we are seeing a combination of (1) an increase in ISM and CGM metallicity at earlier times driven by increased gas heating and loss of total gas mass to other phases, (2) boosts in the stellar metallicity at later times driven by a lack of metal dilution by accreted pristine gas from the IGM and (3) decreases in the total amount of metals in stars and the ISM (and to a smaller extent the IGM) by the decrease in SFRD at $z<6$. 

\subsubsection{Metal absorption equivalent width distributions}\label{ssec:ew}
\begin{figure*}
\includegraphics[width=0.9\textwidth]{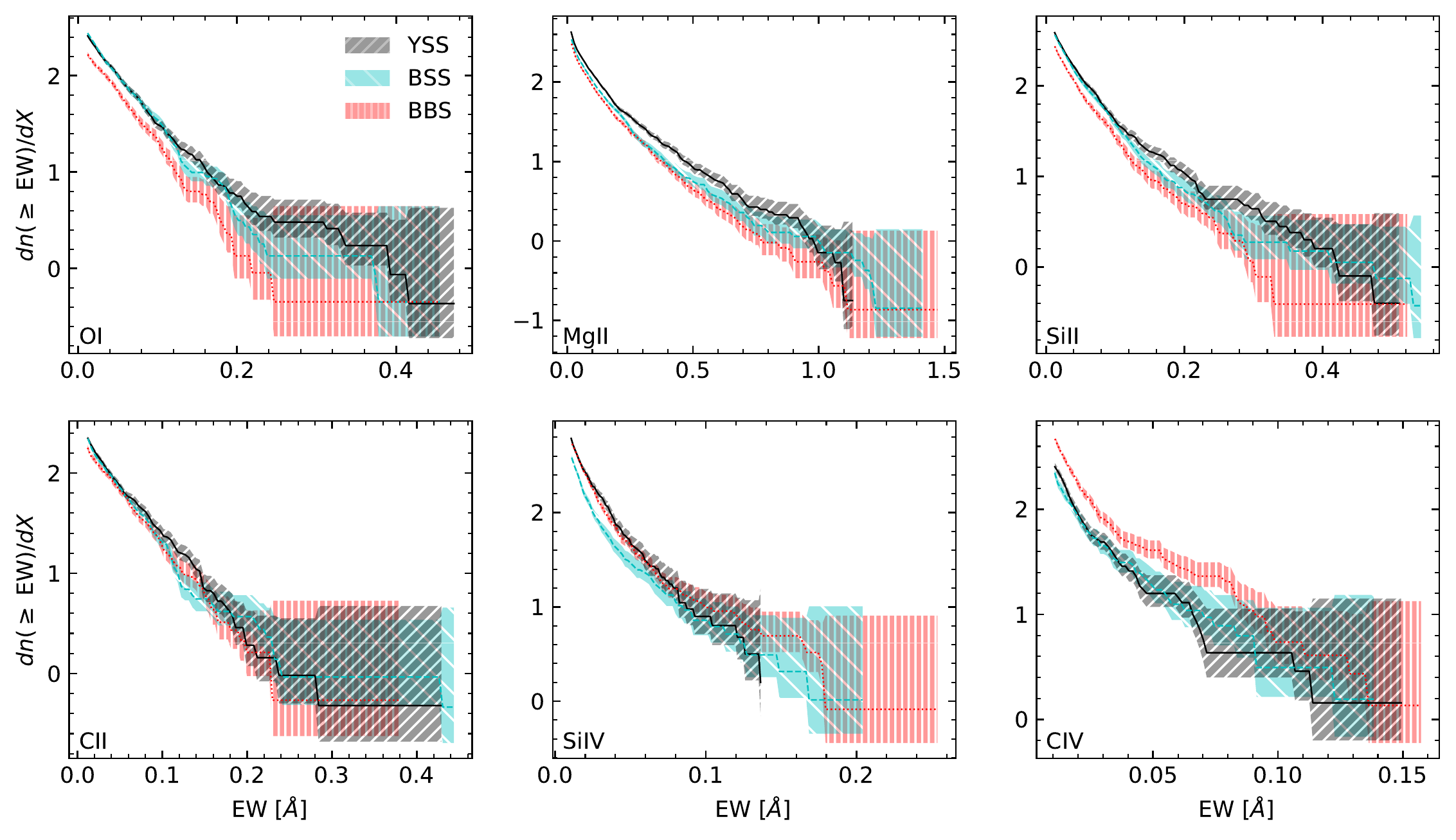}
\caption{For $z=6.5$, the cumulative pathlength-weighted equivalent width distributions of the six tracer ions as modeled in simulations using YSS (black), BSS (cyan), and BBS (red). The error in each EW bin is assumed to be Poissonian, and therefore the bins with small numbers of counts (i.e. those with high EW) have large uncertainties. The solid, dashed, and dotted lines indicate the predicted line incidence in each bin.} \label{fig:EWD}
\end{figure*}

The equivalent width distribution (EWD) is used to describe the intrinsic distribution of strengths of metal absorption systems, and can serve as a metric for testing feedback models and other aspects of simulation implementations~\citep[see e.g.][]{keating16}. It is calculated as the number of absorption systems with a given EW divided by the probed pathlength. For each simulation snapshot, which represents a single redshift, the pathlength $dX$ is calculated according to
\begin{equation}
    dX = \left( 1 + z\right)^3 \frac{dv}{c} \left(\Omega_m \left(1+z\right)^3 + \Omega_\Lambda\right)^{-1/2}
\end{equation}
where $dv$ is the Hubble velocity spanned by the simulated sightline, $4 \times 10^6$ km/s here, and $\Omega_m$ and $\Omega_\Lambda$ are the cosmic density of matter and dark energy, respectively. We plot the simulated cumulative $dn/dX$, where $dn$ is the number with $\mathrm{EW}\geq$ a given EW, at $z=6.5$ in Figure~\ref{fig:EWD}. We select $z=6.5$ because reionization has not yet completed in any of the three models, so the metagalactic UVB has not yet undergone any drastic evolution.

Three of the six tracer ions (O I, Si II, and C IV) roughly follow the trends we expect based on the harder spectrum of the binary stars. The two single star models result in similar distributions for O I and Si II, creating more and stronger systems relative to BBS. C IV, which requires photons with $E=3.5$ Ryd to be created from C III, is most abundant in the harder BBS model, particularly below 0.1 $\angstrom$. Mg II, C II, and Si IV on the other hand deviate from the naive explanation that seems to apply for these other ions. Mg II is slightly more abundant for YSS, particularly at intermediate strengths, but appears at nearly the same rate and strength in the two BPASS models, despite the sizeable differences in their SEDs.  C II is similar for all three models, while for Si IV the distributions are actually closer between YSS and BBS for EW$<0.14 \angstrom$ than between the two implementations of single stars.

For an additional point of comparison, we examine the differential EWDs for O I and Mg II at $z=6.5$ (Figure ~\ref{fig:OI_MgII_EWD}), with overplotted fits from~\citet{becker19} and~\citet{chen17} for O I and Mg II, adjusted to our cosmology. \footnote{We restrict ourselves to O I and Mg II because they are the only ones with observation data available at the correct redshifts, and with a best-fit distribution that includes EW-based completeness corrections.} For O I, there are generally more systems made in the single star models, though there is overlap for $\mathrm{EW} > 0.2 \; \angstrom$. The single star distributions also extend to higher EW than for BBS. The distributions for Mg II are essentially identical for all three models, although the systems reach higher EW in the single star models.

The maximum likelihood fit from~\citet{becker19} is for their O I observations above $z=5.7$, since they observe a strong decrease in the O I incidence rate above $z\sim5.7$. The fit from Mg II from~\citet{chen17} is derived for $z>6$ for the same reason. While the O I incidence found in the observations is matched by the simulations for systems with EW roughly equal to 0.13 $\angstrom$, the slopes of the simulated distributions are too steep and underproduce stronger systems. The exception is in the single star models which, due to low numbers of systems above $\mathrm{EW}\sim0.25 \angstrom$, have quite large errors and therefore overlap with the observations. Following a similar trend, for Mg II all three SPS models significantly underproduce systems of $\mathrm{EW}>0.3 \; \angstrom$, and overproduce weaker systems. Similarly to the O I distributions, higher equivalent widths have a flatter distribution with larger uncertainties.

\subsubsection{Comoving metal absorber incidence rate}\label{ssec:dndx}
To examine the evolution of metal species with time, we plot the incidence rates for the three SPS models against redshift (Figure~\ref{fig:dNdX}). The incidence rate, or line of sight number density of each ion, is the integral of the differential EW distribution function. To make an equivalent comparison to observations and remove the high numbers of weak systems that are not realistically detectable, we apply lower limit cutoffs in either EW or column density, whichever measurement is reported in the corresponding observational comparison. 
Further, while previous studies have demonstrated differing redshift evolution dependent on the strength of the absorber~\citep[although this is only seen for some transitions, see e.g.][]{narayanan07}, we restrict the systems in the figure only with a single lower limit cutoff rather than differentiating between ``strong'' and ``weak'' systems.

\begin{figure*}
\includegraphics[width=0.8\textwidth]{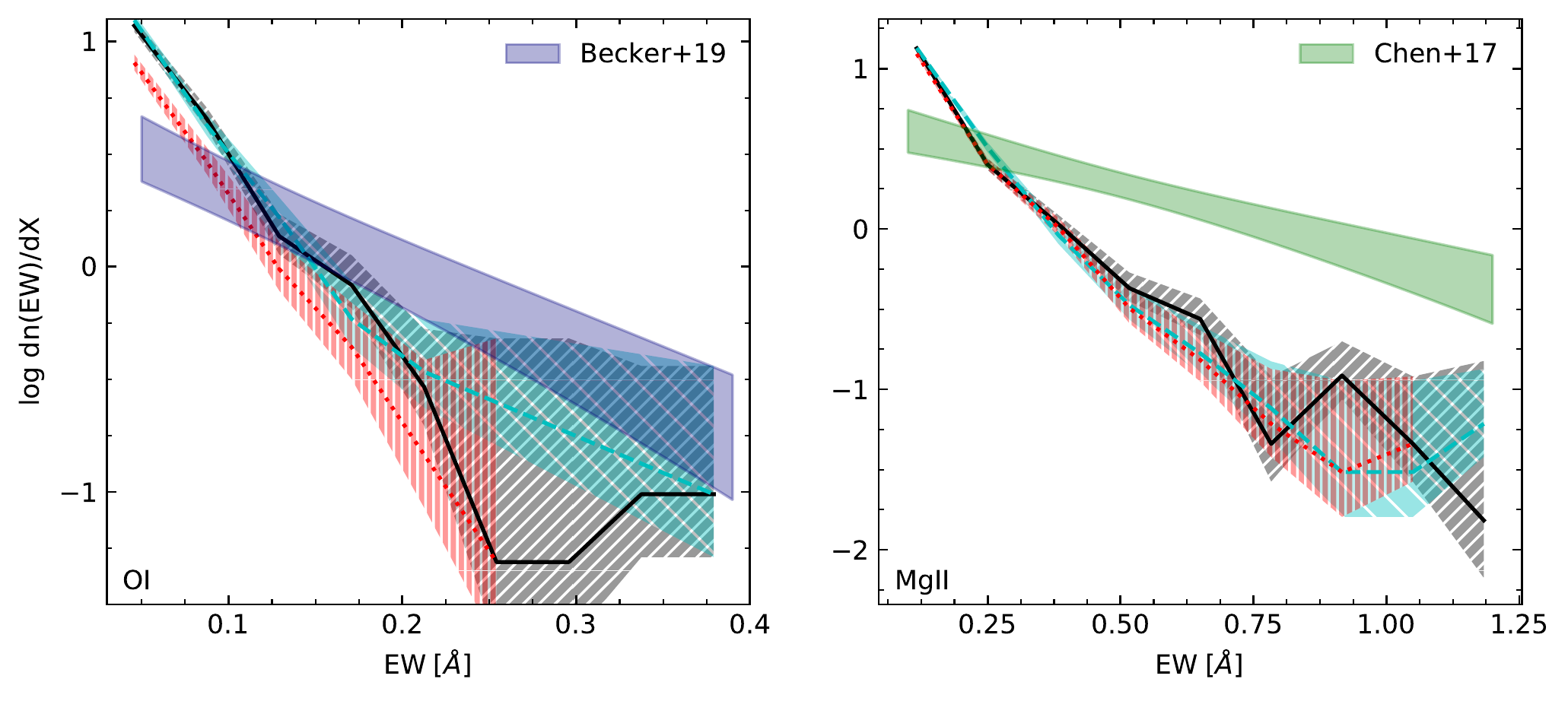}
\caption{The EW distributions of O I (left) and Mg II (right) with overplotted fits from~\citet{becker19} and~\citet{chen17}, which both include completeness corrections. The errors are assumed to be Poissonian and based on the number of counts in each bin. The lines are as in Figure~\ref{fig:EWD}.}\label{fig:OI_MgII_EWD}
\end{figure*}

\begin{figure*}
\includegraphics[width=0.9\textwidth]{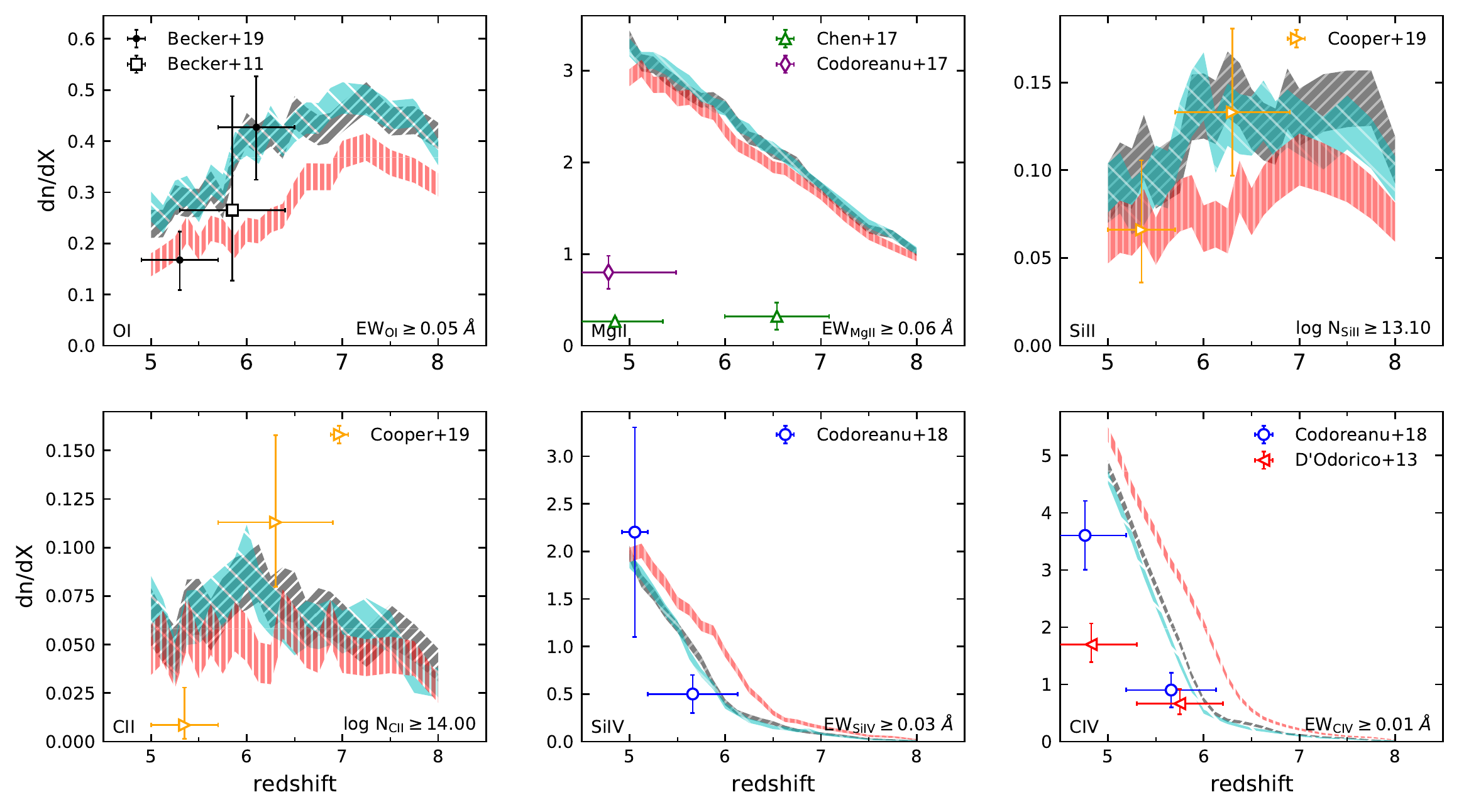}
\caption{Incidence rates of six tracer ions from $z=8$ to 5 for the three SPS models, with some observations overplotted. Reported completeness-corrected incidence rates are overplotted from~\citet{becker19},~\citet{chen17},~\citet{cooper19}, and~\citet{codoreanu18}. The simulated incidences have been calculated using cutoffs set to the minimum value observed in each work, with the exception of~\citet{cooper19}, where the cutoffs are $\log N_\mathrm{CII}=14.0$, where they report near unity completeness, and $\log N_\mathrm{SiII}=13.1$, the equivalent optical opacity for Si II based on the C II cutoff.}\label{fig:dNdX}
\end{figure*}

As redshift decreases, O I, Si II, and C II all undergo a period of growth at higher redshifts before reaching a peak and diminishing; this proceeds regardless of the implemented SPS model, although the peaks in the binary star model are systematically lower. The overplotted observations are from \citet{becker19} (O I) and \citet{cooper19} (Si II, C II). The \citet{becker19} observations, with $\mathrm{EW} \geq 0.05 \angstrom$, show that the $z\sim6.1$ incidence is in good agreement with the single star models, but by $z=5.3$ the simulated incidence of these has not quite diminished enough to match the observations. Instead, the lower incidence of BBS is a better match.  The \citet{cooper19} observations of Si II match the incidence produced by single stars quite well for $z=6.3$ while still overlapping the binary star predictions, but C II only agrees for single stars. The relative insensitivity of Mg II to binary stars seen in Figure~\ref{fig:OI_MgII_EWD} reappears in analysis of the incidence rate, and in contrast with the other low-ionization systems the overall trend is an increase in systems with decreasing redshift. The two single star models show virtually no difference between one another, and BBS creates slightly fewer systems. However, all three models dramatically overproduce the observations of~\citet{chen17}, a difference which is likely driven by the overabundance of weak systems in the simulations.

In contrast to O I, Si II, and C II, the overall Si IV and C IV incidences both show general increases in BBS in the number of systems throughout $z=8$ to 5. Si IV systems $\mathrm{EW} \geq 0.03 \angstrom$ are more abundant in the BBS model, and it outproduces BSS and YSS until $z=5$ when they converge to roughly the same values. Observations taken from~\citet{codoreanu18} at $z\sim5.6$ are consistent with the single star models but are lower than predictions including binary stars. By $z=5$, all models agree with the observations. C IV increases similarly to Si IV, although the rate is much faster, more than a factor of 8 increase from $z=6$ to $z=5$ for single stars. Here, the two single star models have similar results, but BBS consistently has more systems at a given redshift. The BBS C IV incidence also experiences very strong growth, although at slightly earlier times. Compared to the $dn/dX$ reported in~\citet{codoreanu18}, binary stars overpredict C IV incidences while the single star models perform acceptably. The simulations all predict a much higher C IV incidence than seen in~\citet{dodorico13} at $z=5$.

While agreement between the simulated and observed low ionization incidences is quite variable, (for example, several ions agree better at $z\sim6$ with single star models and at $z\sim5$ with binary star models) there is better agreement for the high ions with the single star models. However, this does not necessarily indicate that high binary fractions during the EoR are ruled out. In particular, these high ions appear to be very sensitive to the timing of reionization, which is dependent on our implementation of the escape fraction. We discuss this further at the end of Section~\ref{sssec:disc_metal_absorbers}.

\subsubsection{Comoving metal mass density}\label{ssec:Omega_metal}
\begin{figure*}
\includegraphics[width=0.9\textwidth]{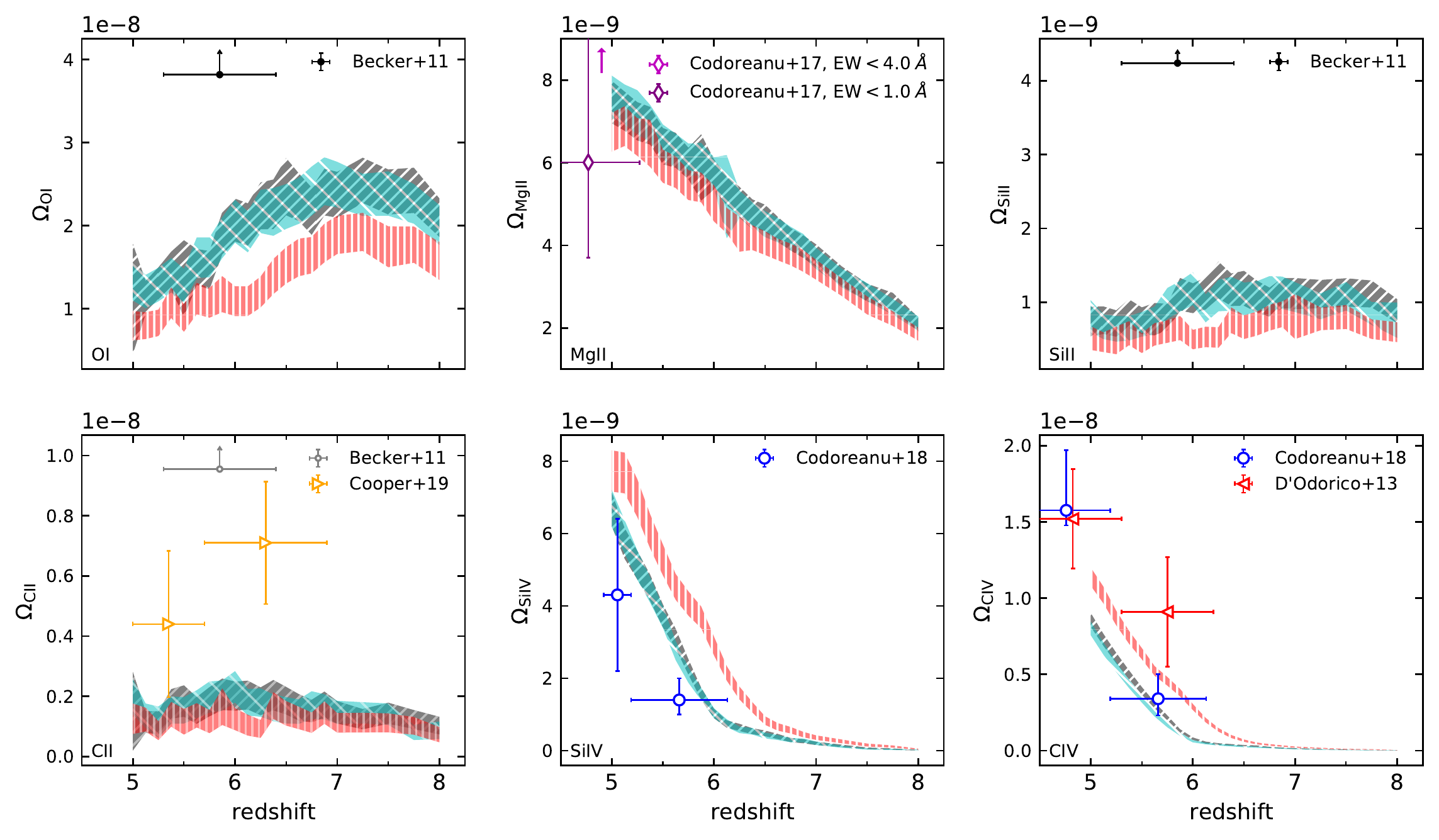}
\caption{Comoving mass density of commonly observed transitions, with observations overplotted. All observations were completeness-corrected in the original works, and we have taken the reported values and corrected for the cosmology differences at the average reported redshift.} \label{fig:comoving_density}
\end{figure*}

The comoving mass density measures the abundance of a particular ion, defined such that any evolution with redshift is evidence of variation independent of the Hubble flow. Given with respect to the present day critical density, it is calculated as
\begin{equation}
    \Omega_\mathrm{ion} = \frac{H_0 m_\mathrm{ion}}{c \rho_\mathrm{crit,0}} \int N f\left(N, X\right) dN
\end{equation}
where $\rho_\mathrm{crit,0}=8.6\times10^{-30} \; \mathrm{g/cm^{-3}}$, $N$ is the column density and $f\left(N,X\right)$ is the column density distribution function, $\equiv \partial^2 n / \partial N / \partial X$. In practice, the integral is often computed as a discrete sum of observed systems' column densities
\begin{equation}
    \int N f\left(N, X\right) dN = \frac{1}{dX} \sum_{i=1}^{N_\mathrm{sys}} N_i
\end{equation}
We replicate this procedure to compare to observational works in Figure~\ref{fig:comoving_density}. The same lower limit cutoffs in column density or equivalent width are applied to the simulations as in subsection~\ref{ssec:dndx}. Since the cosmic densities are calculated directly from the incidence, the evolutionary trends are roughly the same for $\Omega_\mathrm{ion}$ as for $dn/dX$, and so the focus of this figure is the comparison to observations.

\citet{becker19} at $z=5.7$ provides a lower limit on the $\Omega_\mathrm{OI}$, but it is a factor of two or so above the predictions from the single star models, and the discrepancy is larger for the binary model. The total Mg II is underproduced by all three models, with $\Omega_\mathrm{MgII}$ lower by more than an order of magnitude compared to observations of $\mathrm{EW}<4.0 \angstrom$ systems from~\citet{codoreanu17}; at $z=4.77$, they find $\Omega_\mathrm{MgII} = 3.8^{+6.2}_{-2.3} \times 10^{-7}$. In Figure~\ref{fig:EWD} we see that none of the models generate many Mg II systems at $\mathrm{EW}>1.0 \angstrom$, and Figure~\ref{fig:OI_MgII_EWD} also shows that systems above $\mathrm{EW}\sim0.3$ are significantly underpredicted, so the low cosmic density does not seem surprising. Restricting the comparison to their reported numbers for $\mathrm{EW} < 1.0 \angstrom$, the observations are in agreement with all three models at the 68 per cent level. However, since the observations indicate that the vast majority (98 per cent) of Mg II is in systems with $\mathrm{EW} > 1.0 \angstrom$, this is a large point of contention with the simulation, regardless of the SPS model.

The carbon and silicon systems have mixed results. The simulated $\Omega_\mathrm{CII}$ at $z\sim6.0$ in both the single and binary star simulations is underproduced compared to~\citet{cooper19} and~\citet{becker19}, while the single star simulations barely match~\citet{cooper19} by $z=5.3$. It is important to note, however, that the observational results from these two works are also in conflict with one another, with \citet{becker11} presenting a slightly higher result than~\citet{cooper19}. $\Omega_\mathrm{CIV}$ predicted by all stellar models are in good agreement with observations from~\citet{codoreanu18}, showing the strong increases in overall C IV that have been observed in previous works~\citep[e.g.][]{ryan-weber09}. Only $\Omega_\mathrm{CIV}$ from binary stars matches the observations from~\citet{dodorico13}. Compared to observations, Si II is underproduced by all three models at $z=6$ by at least a factor of four, with the two single star models producing the highest mass densities of Si II. However, Si IV is produced in roughly the correct quantities by the single star models.

Generally then, through inclusion of a binary star-dominated ionizing background, we find that the incidence and comoving density of high (low) ions are increased (decreased). The binary stars also cause earlier evolution of the ions that are seemingly most sensitive to the background, in particular O I, Si IV, and C IV, likely because of the earlier H I reionization occurring in this simulation and the corresponding temperature increases in CGM gas. We discuss these and other implications in more depth in Section~\ref{sssec:disc_metal_absorbers}.

\section{Discussion}\label{sec:discussion}
Our analysis examined the effects on star formation, H I and He II reionization, gas phase, and metal absorber statistics during the EoR as predicted by three SPS models. In sum, there is a complicated interplay between star formation, temperature, decreased enrichment, and spectral shape that results in binary stars having higher abundances of both H I-ionizing and He II-ionizing photons, producing an accelerated H I reionization history, decreasing the star formation rate density and thus generation of metals by stars, and elevating the equivalent widths and incidences of high ionization state metal ions when compared to single stars.
Simultaneously, the strength and incidence of low-ionization states are generally reduced.

\subsection{Star formation and galaxies}\label{ssec:disc_star_formation_etc}
In the binary star simulation, we find that the SFRD drops below $z=6$ and reaches 90 per cent of the rate in the single star simulations by $z=5$. There is also a decrease in the UVLF values for $M_\mathrm{UV}>-15$. These seem to be associated with a diminishing gas fraction in galaxies with $\log M_* < 8$, meaning the cause is at least partially a reduction in the gas available for forming stars. ~\citet{rosdahl18} found similar effects of binary stars on the star formation in galaxies. In particular, when binary stars were included the star formation rate and stellar masses for a given halo mass were systematically diminished, and the luminosity function also showed a lower number of galaxies per luminosity bin, although the differences were apparent over a larger luminosity range than seen here.

The decrease in SFRD in BBS is restricted to galaxies with $M_\mathrm{UV}>-15$, meaning that galaxies with smaller stellar masses should be preferentially affected. We find that there is no clear difference in the stellar mass function between models between $z=7$ and $z=5$ for halos above our resolution limit, but there is a decrease in number for lower masses. Additionally, although observations of the UVLF exist for $z>5$, these studies have not been sensitive enough to very faint galaxies to offer any constraining power yet. Future observations may yet probe to $M_\mathrm{UV}>-15$, but small uncertainties would be required.

\subsection{H I Reionization}\label{ssec:disc_HI_reionization}
The YSS and BSS models lead to identical histories in H I reionization, with the volume-weighted neutral hydrogen fraction reaching $\sim$ 0 at $z=6.0$. The photoionization rates due to galaxy and quasar contributions are also identical, with abrupt increases in the rate occurring as reionization is completed. In comparison, BBS yields consistently higher photoionization rates, which allows reionization to complete sooner, at $z\approx6.3$. The variations in photoionization in response to reionization, particularly for the quasar contributions which can only be affected here by changes in IGM opacity, also occur more quickly for BBS, suggesting that there may be differences in the time scale for homogenization of the opacity of the IGM.

These results are in qualitative agreement with other studies.~\citet{rosdahl18} found that inclusion of binary stars in the {\sc sphinx} code can complete H I reionization by $z\sim7$, while a simulation with single stars is only 50 per cent complete at $z=6$. This contrasts with our results for BSS and YSS, as well as those of~\citet{stanway16}, where both models result in a reionization history that approximately matches the observational constraints. This is likely due to the differing treatments of the escape fraction of galaxies; while in the fiducial TD$+$YSS used in our previous works, the escape fraction has been tuned to cause reionization to end at $z=6$, {\sc sphinx} includes no tuned escape fraction, simply setting it to unity for physical scales below their maximum spatial resolution.

Given the decrease in the star formation rate in BBS we may have expected to see a decrease in the photoionization rate of H I and He II as the SFRD dropped markedly at $z\sim6.0$. BBS does have a lower slope proceeding to lower redshifts in $\Gamma_\mathrm{HI}$ and $\Gamma_\mathrm{HeII}$ for redshifts below 6, but the differences visually appear quite small. Additionally, the three models only begin to significantly diverge at $z<6$, after H I reionization is complete.

\citet{stanway16} determined that ionizing flux is increased substantially for stellar metallicities below 0.3 $Z_\odot$, although this decreases for stars at $\sim$solar metallicity, and~\citet{rosdahl18} similarly saw that binary stars raised the total luminosity and increased the escape fraction by a factor of three, the latter due to binary population luminosity declining at a slower rate with age than for single stars. Since the $f_\mathrm{esc}$ model in TD is only redshift-dependent, we make a simple analytical comparison between the {\sc bpass} single and binary star models using the ratios of their SED amplitudes within the H I-ionizing cross-section. We consider as a test case a 100 Myr-old stellar population with a constant SFR of $1 \; M_\odot$ per year and  a metal mass fraction of $\log Z/Z_\odot= -1.15$ (selected to match the median mass-weighted value in our simulations at $z=6$, which is $\sim -1.2$). We find there is an increase of 36 per cent in the instantaneous number of H I-ionizing photons for binary stars compared to single stars.~\citet{rosdahl18} used ray tracing in post-processing to ascertain escape fractions for resolved structures, so the greater increase they report is certainly more accurate than the highly simplified calculation we make above.

\citet{gotberg20} studied effects of massive blue stragglers and stripped binary stars, both byproducts of binary star evolution, on reionization of H I and He II through use of their stripped star spectral models and a semi-analytical model for reionization. They determined that stripped stars alone may contribute over twenty per cent of the photons contributing to H I reionization.~\citet{secunda20} similarly found higher escape fractions using high resolution cosmological simulations with radiative transfer~\citep{kimm14}, with a stronger effect arising from the stripped stars. {\sc bpass} includes models for both stripped stars and blue stragglers, so their effects are included in our results, although the models are different than the implementations in these other works.

To consider another metric of reionization, we calculate the mean transmission in Ly$\alpha$ for the two {\sc bpass} models. Binary stars create a $T_\mathrm{Ly\alpha}$ of 0.050 (0.203) at $z=6$ ($z=5$) compared to 0.002 (0.123) in BSS. Observations have settled on $T_\mathrm{Ly\alpha} \approx$ 0.006 (0.13) at $z=6$ and 5 respectively, so the binary star emissivities for a given redshift are certainly too high, indicating that our BBS simulation is probably not accurate. However, again, the fiducial TD with {\sc yggdrasil} is tuned via the sub-grid escape fraction model such that reionization completes near $z\sim6$. Therefore, if the simulation were re-tuned to accommodate the BBS model such that reionization concluded later, it is possible that the predicted metrics would be more in line with the observations. This means that we can rule out our fiducial escape fraction model in combination with binary stars, though not the presence of binary stars themselves.

\subsection{He II reionization and the IGM temperature at mean density}\label{ssec:disc_HeII_reionization_IGM_T0}
We found little difference in the expected He II reionization history regardless of the SPS model used in TD, with each of the three models resulting in $x_\mathrm{HeII,v}$ and $x_\mathrm{HeIII,v}$ of approximately 50 per cent by $z=5$. When EoR galaxy emissivities include significant contributions from binary stars, quasars become dominant in the He II-reionization process at later times, by tens of millions of years, due to the binary stars' higher overall He II-ionizing emissivities. However, the excess heating they cause in the CGM and subsequent decrease in star formation rate leads to slightly slower rates of He II$\rightarrow$He III ionization at later times, as the single stars showed slightly higher $x_\mathrm{HeIII,v}$ by $z=5$.

Our ability to make more detailed predictions regarding He II reionization is limited because we have restricted ourselves to studying outputs at $z \geq 5$. He II reionization is believed to have completed at $z\sim3$ due to the measured quasar emissivity being insufficient at earlier redshifts~\citep{masters12,furlanetto08,faucher_giguere09,hm12}. However, early He II reionization can be indirectly studied through its effects on the temperature at the mean density of the IGM. With the inclusion of multi-frequency radiative transfer in TD, the trajectory of the evolution in $T_0$ is essentially the same for all models, with a gradual increase with decreasing redshift until the conclusion of H I reionization, at which the temperature peaks and then begins to drop before beginning to rise again. Due to the earlier reionization produced by BBS, this peak temperature is reached at an earlier time, however, the value of the peak temperature is virtually identical in all three models, and there is only a small temporal offset between the trends. The binary stars do result in an increase in the temperature at mean density of $\sim$800 K at redshifts prior to completion of H I reionization, but the situation reverses at lower redshifts, with single stars resulting in a slightly higher $T_0$. ~\citet{gotberg20} found the IGM temperature increased by 1000-5000 K when including binary stars in their analytical models of H I reionization, which is higher than in our results.

While the general trend of the IGM temperature evolution is likely correct, the simulations produce mixed quantitative results when compared to observations. For example,~\citet{gaikwad20} inferred $T_0$ values near $z\sim5.6$ of about $1.1 \times 10^4$ K, which agree with all three models, and simultaneously agree with estimates from~\citet{puchwein19} and~\citet{upton_sanderbeck20}. However,~\citet{boera19} and~\citet{walther19} arrived at markedly lower values, $\sim 0.6 \times 10^4$ K, out of agreement with~\citet{gaikwad20} and~\citet{puchwein19} at $z=5.4$, before both show increases at $z\leq5$. The discrepancy between observations could perhaps be a side effect of measurement techniques.~\citet{boera19} and~\citet{walther19} arrived at their measurements of $T_0$ using the transmitted flux power spectrum (FPS), which measures the pixel clustering in flux transmission spikes. In contrast,~\citet{gaikwad20} fit observations of Ly$\alpha$ forest transmission spikes using Voigt profile parameters, in tandem with results from the Sherwood-Relics simulation suites~\citep[based on the Sherwood suite described in][]{bolton17}.~\citet{gaikwad20} attributes the discrepancy between these $z=5.4$ inferences to the sensitivity of the FPS to the continuum fitting and uncertainties in the H I photoionization rate, and demonstrates that the spike width distribution they use is not as sensitive to these variables.

~\citet{puchwein19} uses analytical models of the ultraviolet background (UVB) on cosmological scales with updated ionizing emissivities from stars and AGN, and new prescriptions for treating reionization of the IGM. Their fiducial model creates decreasing evolution of $T_0$ with decreasing redshift that is steeper than in TD, although the inferences from~\citet{gaikwad20} are in agreement with their model.~\citet{upton_sanderbeck20} focuses on He II reionization using a model to reproduce the physical effects of the process while avoiding full \textit{in situ} radiative transfer, and has a fiducial UVB model that matches~\citet{puchwein19} above $z=4$. Thus, since there remains this consistent disagreement between FPS-derived $T_0$ and the values predicted by both numerical simulations and analytical models, it is difficult to determine whether our models are consistent with these other works. We can say, however, that all three SPS model results lie within the ranges predicted by these other studies when considered as a whole.

\subsection{The circumgalactic medium and metal absorption systems}\label{ssec:disc_metal_absorbers}

\subsubsection{CGM temperature}
Binary stars result in an increase of the CGM temperature of $\sim$3000 K on average across $z=5$\textendash 8, for both the dense and diffuse CGM phases. Further, this increase in temperature is sustained, albeit diminished, even into the post-reionization regime. This contrasts with the results for the IGM, where the temperatures for all models converge at later times. By $z=6$, through additional heating, binary stars have removed $>$10 per cent of the gas from the CGM overdensity regime ($1 < \log \Delta < 2.5$) and deposited it in the IGM, and similarly removed $>10$ percent of material from the ISM. Since star formation is contingent on the presence of cold, dense gas, this increase in temperature of denser gas within the binary star simulation is adversely affecting the star formation rate. Additionally, from this general increase in the CGM temperature (as well as the established increase in spectral hardness), we predict that there should be variation in the ionization states of metals observed around EoR galaxies dependent on their stellar population binary fractions.

\subsubsection{Metal absorption systems}\label{sssec:disc_metal_absorbers}
The two single star models create fairly similar cosmic densities of metal absorption systems despite the differences in their SEDs. Compared to the {\sc bpass} binary star model, they result in higher (lower) incidences of low (high) ionization states of metals. Other works have similarly determined that binary star backgrounds should elevate high ionization states such as C IV and Si IV~\citep{gotberg19}.

The incidence rates of O I, Si II, and C II from all models approximately match the observations from~\citet{becker11},~\citet{becker19}, and~\citet{cooper19}. However, predictions from the single star models generally produce better overlap between the observations, with the exception of O I at $z=5.4$. In contrast to this good performance, the low ionization ions have consistent difficulty matching the observed comoving densities. $\Omega_\mathrm{ion}$ in O I, Si II, and C II from both the single and binary star models disagree with measurements from~\citet{becker11}. C II produced by single stars at $z=5.3$ matches~\citet{cooper19}, but underpredicts their values from $z=6.3$.

We generally see with the low ions that even if $dn/dX$ approximately matches the observations, the total abundance is still far too low. While the incidence is a very simple metric, the abundance has more dependencies, and in practice the $\Omega_\mathrm{ion}$ inferred from observations is dominated by high column density systems. Since our differential EWD plots of shows a tendency for TD to overpredict weak systems and underpredict strong ones we can see that this is a likely source of the disagreement.

Mg II behaves uniquely here, as there are only minor differences between the incidences and comoving densities in the single and binary  star simulations. More importantly, irrespective of the choice of SPS model the predicted Mg II EWD is too steep, drastically overproducing systems with EW $<$ 0.2 $\angstrom$ at $z=6.5$ and failing to generate the strong systems that dominate the comoving density at all redshifts. Additionally, the abundance of weak systems grows significantly in the post-reionization epoch, in contrast with the observed relatively flat evolution~\citep{chen17}.

In contrast to the low ionization ions, the high ionization states do a fairly good job of matching both the incidences and comoving densities seen in observations, although the agreement varies a bit with the SPS model and the redshift. Generally, the single star model predictions are closer to the observed values, but one exception is in the $z=5.6$ comoving density of C IV. Here, the observations from~\citet{dodorico13} 
and~\citet{codoreanu18} disagree, with~\citet{dodorico13} being a factor of $\sim$2 higher. In this case, only the binary stars match the observations. We have previously highlighted an underproduction of strong C IV in the cumulative EWD by TD compared to~\citet{dodorico13}~\citep[see][]{finlator20}, but given its discrepancy with~\citet{codoreanu18}, it may not be as critical an issue as was previously asserted.

From the underproduction of low ionization ions, there is a general indication that there is too little gas at the right temperature and/or density to accommodate the low ionization states. It further seems to be a systematic discrepancy between these simulations and observations, and not a result of the particular SPS model in use. As mentioned in subsection~\ref{ssec:disc_HI_reionization}, since $T_\mathrm{Ly\alpha}$ in BBS is too high for a given redshift compared to observations, it is likely that for this model there are too many ionizing photons contributing to low incidences of the low ionization states. However, this is not the case for the single star models, so there are factors distinct from the UVB playing a role.

The relatively small simulation volume could provide an explanation for the discrepancies seen between observations and the simulated metals, at least for some ions. The small volume results in only a subset of the true galaxy population at $z\geq5$, with TD's simulated stellar masses only reaching $\log M_* = 8.8$ at $z=6$, while observations detect galaxies up to $\log M_* = 11$~\citep{furtak21}. Thus, if strong absorption systems are associated with higher stellar mass galaxies then the dearth of strong absorbers in our simulations would be expected. This may be the case for Mg II, since its strength seems to be correlated with galaxy stellar mass for blue galaxies at lower redshifts~\citep{bordoloi11}. However, other simulations with larger box sizes also show difficulty reproducing strong Mg II~\citep[e.g.][]{keating16}. The underlying physical prescriptions of these simulations are quite different though, so this is still an important avenue to explore in the context of TD. On the other hand, C IV strength may be correlated with sSFR~\citep{bordoloi14}, which is in turn inversely correlated with $M_*$. If the EW$(\mathrm{CIV})$-sSFR correlation is real, TD should in theory be capable of creating strong C IV.

Another possible factor in the discrepancies could be the simulation resolution. Here, $M_\mathrm{gas} \sim 1 \times 10^6 \; M_\odot$ which, while fairly high for a cosmological simulation, may still not be high enough to accurately capture the creation and evolution of dense gas within the CGM. Higher spatial and mass resolutions in zoom-in simulations are seen to create larger quantities of cool, dense gas structures in the CGM, as a consequence increasing the abundance of low ionization ions~\citep[see e.g.][]{hummels19,vandevoort19}. Other studies indicate that absorber characteristics derived from simulated spectra of both low and high-ionization ions can be affected by resolution~\citep{peeples19}. In~\citet{finlator20}, we tested for convergence of the C IV and Si IV column density distribution functions and found that $M_g \leq 5 \times 10^5 \; M_\odot$ was required for systems at $\log \; N > 12$ to converge; thus, we expect that the statistics of Si IV and C IV should nearly be converged in these simulations. However, it is still conceivable that implementing an increased mass resolution may boost the strengths of our low ion absorbers.

~\citet{becker19} observed a pronounced decrease of a factor of $\sim2-4$ in the incidence of O I absorption systems across $z=5.7$ when binning systems in sizes $\Delta z = 0.8$, but shrinking the bin size to $\Delta z = 0.4$ suggests the decrease could be underway as early at $z=6.1$, albeit with high uncertainty. Our previous investigations of the evolution in O I during reionization, using our fiducial {\sc yggdrasil} single star SPS model, showed a roughly level distribution of O I with $\mathrm{EW} \geq 0.01$ from $z=8\rightarrow6.5$ that diminishes as reionization is completing~\citep{doughty19}. However, this work uses a different version of TD~\citep[see][for details]{finlator18,finlator20}. There have also been improvements made to our spectrum-generating code, including changes to the velocity range over which particles are allowed to contribute to the optical depth for a given transition. All three models now show diminishing values from higher redshifts, up to $z=7$, which better reproduces the trend seen in the $\Delta z = 0.4$ bin sizes in~\citet{becker19}.

\citet{cooper19}, further verifying results of previous works~\citep[e.g.][]{ryan-weber09}, found an abundance of low-ionization  metal systems at $z>5.7$ containing no C IV or Si IV, despite the frequent  occurrence of high ionization states at lower redshifts. Specifically, more than 50 per cent of their detected absorption systems above $z=5.7$ contain only C II, whereas it drops to  20 per cent at lower redshifts. In the same redshift bins, systems containing carbon exclusively in the C IV state increase from 23 per cent to 43 per cent. Pursuing the cause of this dropoff, they question whether it stems from lower levels of enrichment, a softer ionizing background, or weaker galactic feedback at $z>5$.

In our simulations, at $z=6$ a C II-only occurrence is found in 40 (11) per cent and C IV is in 51 (83) per cent of systems in the single (binary) models. By $z=5$, C II is in single digit percentages for both models, and C IV-only appears in 83 and 90 per cent for the single and binary models, respectively. Generally then, while the simulation recreates the strong evolution of C II and C IV absorption, both SPS models produce a high occurrence of C IV-only systems earlier than indicated in the observations. The softer single star SPS model predicts occurrences which are in the best alignment with observations.

In this work, our study of $dn\left(\mathrm{CIV}\right)/dX$ from the single and binary star simulations, which have comparable levels of enrichment down to $z=6$ and identical feedback models, clearly show that the incidence is \textit{strongly} tied to the completion of reionization. The rate of change in incidence increases abruptly once the EoR is completed within the simulation, and the same effect is seen in Si IV.

We perform a simple test to explore this further, isolating the changes in C IV incidence in the {\sc bpass} single star simulation due to the UVB. We accomplish this by using the $z=6.5$ and $z=5.5$ snapshots and comparing the evolution of C IV incidence when spectra are generated using the predicted UVBs from each snapshot versus using the $z=6.5$ UVB in both (these redshifts being selected to bookend the end of reionization). The application of the same UVB to both snapshots isolates the effects of changes in enrichment and other physical conditions from those of the instantaneous UVB. For reference, the background at $z=5.5$ has an amplitude at $E=3.5$ Ryd, $\sim$4 times higher than at $z=6.5$. When the UVB is held constant, the C IV incidence increases by a factor of 2.9 as opposed to 6 when it is changed. Since with the correct UVB the evolution is a factor of 2.1 greater, it seems that enrichment and other physical changes are also very significant. However, taking into account the obvious relationship between the end of the EoR and the C IV incidence, we speculate that the change in C IV incidence is primarily driven by the increasing temperature of the CGM gas in response to reionization.

\section{Conclusions}\label{sec:conclusions}
In this work, we evaluate the effects of the spectral energy distributions characteristic of high binary fraction stellar populations on subsequent galaxy evolution during reionization. We use cosmological simulations which include in-situ multi-frequency radiative transfer to capture the impact of the local ionizing background on the CGM, itself affecting the baryon cycle of galaxies. Our main conclusions are that, all other variables held constant, a high fraction of binary stars:
\begin{itemize}
\item Increases global photoionization rates of H I at all redshifts and of He II at $z>6.3$;
\item Contributes greater photoheating primarily to the CGM causing it to become more diffuse, and reduces typical galaxy gas masses;
\item Reduces star formation in galaxies with $M_\mathrm{UV}>-15$, causing a 10 per cent decrease in the SFRD by $z=5$, which further leads to diminished metal masses in gas below $z=6$;
\item Produces more highly-ionized metal absorption for a given redshift and earlier strong evolution in the incidence of Si IV and C IV systems;
\item Produces metal ion incidences/comoving densities and an H I reionization history that are not preferred over single stars when compared to these observables, at least for our current escape fraction model.
\end{itemize}
To synthesize these, we find that while at very early times the harder spectra and higher effective escape fractions associated with binary stars cause elevated photoionization rates of hydrogen and greater incidence of high-ionization metal ions, these also reduce star formation at later times and by extension the subsequent photon budget. We conclude then that the effects of binary stars are complex and self-regulating due to their own feedback. While our simulations with binary stars do not match observables as well as single stars, this is due to the precise tuning of the fiducial single star simulation, which is set to complete H I reionization by $z=6$. Future work in this area would benefit by simulating to a lower redshift, re-tuning the escape fraction model to force the binary star simulation to reionize later, and incorporating the differing metal yields and SNe rates associated with binary star evolution.

\section*{Acknowledgements}

CCD thanks the LSSTC Data Science Fellowship Program, which is funded by LSSTC, NSF Cybertraining Grant \#1829740, the Brinson Foundation, and the Moore Foundation; Their participation in the program has benefited this work. CCD acknowledges funding from the New Mexico Space Grant Consortium Grant \#NNX15AL51H. This research was supported by the National Science Foundation under Grant No.\ 2006550. Our work made use of the WebPlotDigitizer tool (https://automeris.io/WebPlotDigitizer), for which we thank A. Rohatgi.

\section*{Data Availability}
The data will be shared on reasonable request to the corresponding author.





\bibliographystyle{mnras}
\bibliography{binstars.bib}

\bsp	
\label{lastpage}
\end{document}